%% file: 0_manuscript.tex
\documentclass[manuscript]{acmart}

\usepackage{url}
\usepackage{wrapfig}
\usepackage{graphicx}
\usepackage{color}
\usepackage{multirow}
\usepackage{rotating}
\usepackage{pifont}
\usepackage{acronym}
\usepackage{xcolor}
\usepackage[toc,page]{appendix}
\usepackage{makecell}
\usepackage{subcaption}  
\usepackage{float} 

\newcommand\xmark{\ding{56}}
\newcommand{\cmark}{\ding{52}}

\AtBeginDocument{%
  }

\setcopyright{acmlicensed}
\copyrightyear{2025}
\acmYear{2025}

\makeatletter
\renewcommand{\copyrightpermissionfootnoterule}{%
  \kern-3pt
  \color{white}\hrule width \columnwidth
  \kern 2.6pt
  \color{black}
}
\makeatother

\makeatletter
\def\@mkauthorsaddresses{%
  \bgroup
    \color{white} 
    \def\streetaddress##1{}%
    \def\postcode##1{}%
    \def\position##1{}%
    \gdef\@ACM@institution@separator{, }%
    \def\institution##1{\unskip\@ACM@institution@separator ##1\gdef\@ACM@institution@separator{ and }}%
    \def\city##1{\unskip, ##1}%
    \def\state##1{\unskip, ##1}%
    \renewcommand\department[2][0]{\unskip\@addpunct, ##2}%
    \def\country##1{\unskip, ##1}%
    \def\and{\unskip; \gdef\@ACM@institution@separator{, }}%
    \def\@author##1{##1}%
    \def\email##1##2{\unskip, \nolinkurl{##2}}%
    \addresses\\%
  \egroup}
\makeatother

\usepackage{xcolor} 

\makeatletter
\providecommand\@formatdoi[1]{\textcolor{white}{https://doi.org/#1}}
\makeatother

\begin{document}

~\vspace{1mm}
\title{An End-to-End Pipeline Perspective on Video Streaming in Best-Effort Networks: A Survey and Tutorial}

\author{Leonardo Peroni}
\email{leonardo.peroni@imdea.org}
\affiliation{%
  \institution{UC3M}
  \streetaddress{Avenida de la Universidad, 30}
  \city{Leganes}
  \state{Madrid}
  \country{Spain}
}
\author{Sergey Gorinsky}
\email{sergey.gorinsky@imdea.org}
\affiliation{%
  \institution{IMDEA Networks Institute}
  \streetaddress{Avenida Mar Mediterraneo, 22}
  \city{Leganes, Madrid}
  \country{Spain.}}

\renewcommand{\shortauthors}{Leonardo Peroni and Sergey Gorinsky}

\begin{abstract}

\textcolor{black}{
Remaining a dominant force in Internet traffic, video streaming captivates end users, service providers, and researchers. This paper takes a pragmatic approach to reviewing recent advances in the field by focusing on the prevalent streaming paradigm that involves delivering long-form two-dimensional videos over the best-effort Internet with client-side adaptive bitrate (ABR) algorithms and assistance from content delivery networks (CDNs). To enhance accessibility, we supplement the survey with tutorial material. Unlike existing surveys that offer fragmented views, our work provides a holistic perspective on the entire end-to-end streaming pipeline, from video capture by a camera-equipped device to playback by the end user. Our novel perspective covers the ingestion, processing, and distribution stages of the pipeline and addresses key challenges such as video compression, upload, transcoding, ABR algorithms, CDN support, and quality of experience. We review over 200 papers and classify streaming designs by problem-solving methodology, whether based on intuition, theory, or machine learning. The survey further refines these methodology-based categories and characterizes each design by additional traits such as compatible codecs. We connect the reviewed research to real-world applications by discussing the practices of commercial streaming platforms. Finally, the survey highlights prominent current trends and outlines future directions in video streaming.
}
\end{abstract}




\maketitle



\newpage
\textcolor{black}{\section{Introduction}\label{sec:Introduction}}

\textcolor{black}{Video streaming fuels dramatic Internet traffic growth and continues to expand. Video traffic quadruples between 2017 and 2022, increasing its share of total Internet traffic from 75\% to 82\%, while live streaming traffic rises 15-fold~\cite{Cisco2019}. Streaming time grows significantly, with a 90\% increase in Asia and a  14\% global rise between 2021 and 2022~\cite{CONVIVA2022}}. 

\textcolor{black}{Streaming operates in an economically diverse ecosystem.  \textcolor{black}{To boost subscriptions and revenue, \textit{\textsf{streaming platforms}} enhance the \textit{\textsf{quality of experience (QoE)}} for \textit{\textsf{users}}.} \textit{\textsf{Content providers (CPs)}} supply videos, while \textit{\textsf{content delivery networks (CDNs)}} distribute them with low latency, minimizing costs and maintaining high cache hit rates. \textit{\textsf{Internet service providers (ISPs)}} offer network connectivity, characterized by \textit{\textsf{quality of service (QoS)}} metrics. Entities often play multiple roles, and their relationships evolve continuously.}

\textcolor{black}{The technological landscape of video streaming is also heterogeneous and evolving. Major streaming platforms typically exploit the \textit{\textsf{hypertext transfer protocol (HTTP)}}
and scalably distribute video content to global audiences via CDN-assisted \textit{\textsf{HTTP adaptive streaming (HAS)}} where client-side \textit{\textsf{adaptive bitrate (ABR)}} algorithms tackle the diversity and variability of network connectivity between servers and clients. On the other hand, interactive video applications in their real-time communications commonly turn to \textit{\textsf{peer-to-peer (P2P)}} technologies and incorporate their own \textit{\textsf{congestion control (CC)}} algorithms to deal with dynamic network conditions. Even within the HAS paradigm, short-form and 360-degree videos employ  somewhat different streaming techniques than long-form \textit{\textsf{two-dimensional (2D)}} videos. \textit{\textsf{Software-defined networking (SDN)}} and \textit{\textsf{named data networking (NDN)}} represent enhancements of the current Internet architecture that introduce significant new opportunities and challenges for video streaming. \textcolor{black}{Appendices~A and~B expand all mentioned acronyms and offer a glossary of key terms, respectively.}}

\textcolor{black}{This survey provides an extensive overview of recent research on HAS of 2D videos over best-effort networks with client-side ABR algorithms, which constitutes the major paradigm for video streaming on the current Internet. We primarily focus on long-form 2D videos, even though many of the reviewed designs are also relevant to short-form and 360-degree videos. Despite restricting the scope, our survey covers a 
vast amount of material by reviewing more than 200 papers. The survey also includes essential tutorials to make the content accessible, especially for newcomers.}

\begin{figure}[t!]
	\centerline{\includegraphics[trim=1cm 7.5cm 2cm 3cm,clip,width=\linewidth]{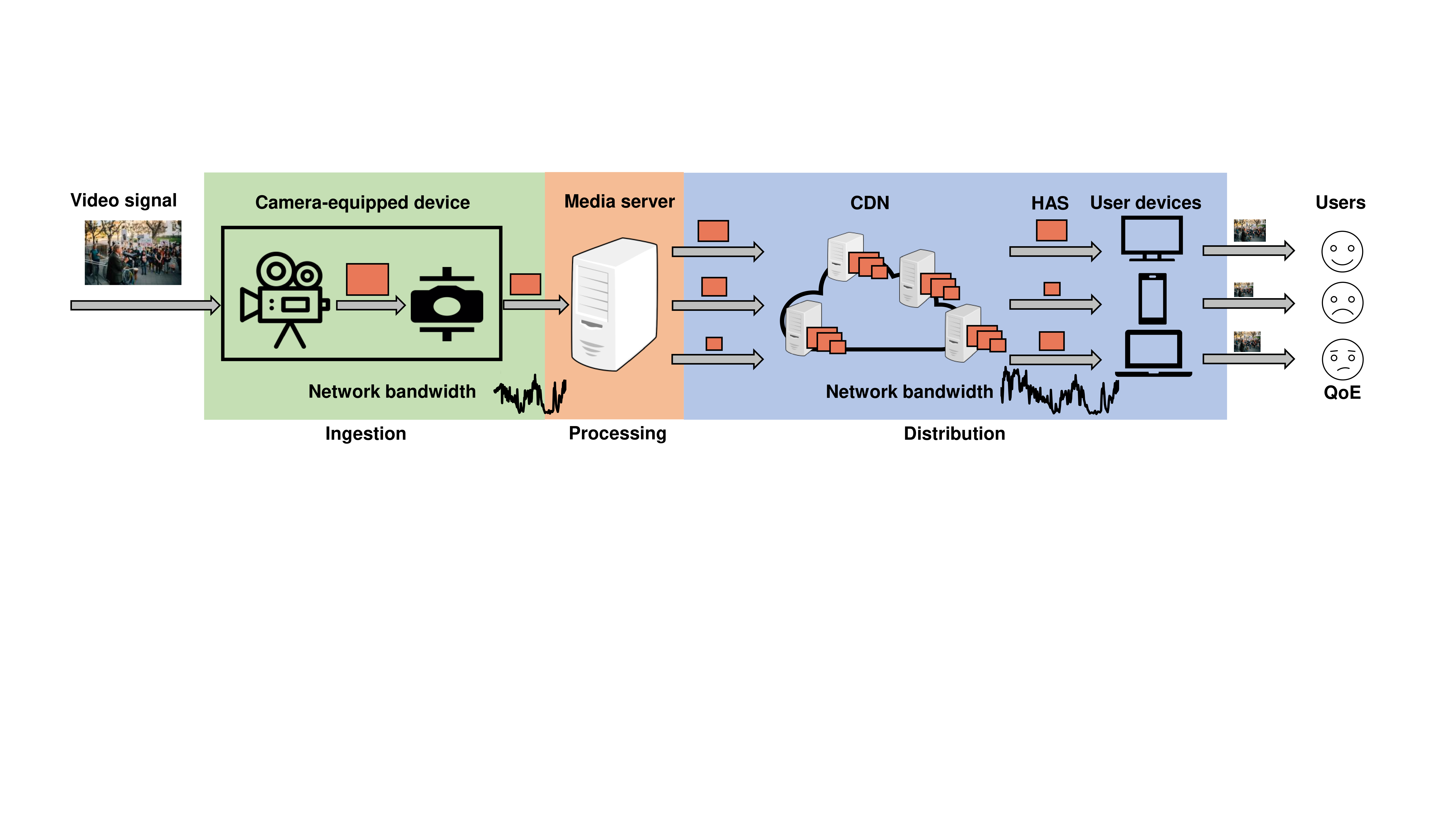}}
	\captionsetup{justification=centering}
 \vspace{-9mm}
	\caption{\textcolor{black}{The end-to-end streaming pipeline and its ingestion, processing, and distribution stages.}}
	\label{end_to_end_pipeline}
  \vspace{-1mm}
\end{figure}

\textcolor{black}{
A major novelty of our approach is in surveying the 2D HAS process from the perspective of its \textit{\textsf{end-to-end pipeline}}. Figure~\ref{end_to_end_pipeline} depicts this pipeline as consisting of ingestion, processing, and distribution stages. At ingestion, a camera-equipped device captures raw footage, encodes it to reduce size, and uploads the video to a media server. The processing stage includes video storage, segmentation, and transcoding to create multiple representations of the video in accordance with an \textit{\textsf{encoding ladder}}. During distribution, a CDN scalably disseminates the video to heterogeneous user devices for decoding and playback. Unlike earlier surveys that focus on individual stages or tasks, we offer an integrated understanding of the entire pipeline. While some techniques and metrics span multiple stages, our survey highlights these interconnections to further promote the holistic understanding.
}

\textcolor{black}{
We introduce a new taxonomy in Section~\ref{sec:Methodologies} and apply it later to structure our discussion of recent works. The stage of the end-to-end streaming pipeline constitutes the top level of the taxonomy, differentiating between the ingestion (Section~\ref{sec:Ingestion}), processing (Section~\ref{sec:Processing}), and distribution (Section~\ref{sec:Distribution}) stages. At the distribution stage, we separately consider its ABR, CDN, and QoE aspects. The lower classification levels categorize each work according to its methodology for tackling the problem. Each leaf category lists its designs in chronological order. In addition to classifying the works, we also describe each of them in terms of various characteristics, such as the codec used. After the thorough review of the research works, we discuss real-world applications
(Section~\ref{sec:case_studies}), trends, and future directions
(Section~\ref{sec:trends_and_future}). 
}
\textcolor{black}{Our survey makes the following main contributions:
\begin{itemize}
\item We present an extensive review of recent research on video streaming and, in particular, CDN-assisted HAS of long-form 2D videos over the best-effort Internet with
client-side ABR algorithms.
\item We cover the topic from the novel perspective of the end-to-end streaming pipeline and discuss more than 200 papers according to a new taxonomy. Within each of the pipeline's ingestion, processing, and distribution stages, the scheme organizes the discussed works based on their problem-solving methodology.
\item Beyond the literature review, we report on real-world applications, trends, and promising future directions.
\end{itemize}
}

~
\vspace{-9mm}
\textcolor{black}{\section{Background}\label{sec:background}}

~
\vspace{-5mm}
\textcolor{black}{\subsection{End-to-end streaming pipeline}\label{subsec:end_to_end}}

The end-to-end streaming pipeline starts at the \textit{\textsf{ingestion stage}} with the capture of raw video by a camera-equipped device, with a codec applying spatial and temporal compression to the raw footage for reducing the video size, and subsequent upload of the encoded video over the Internet to a media server. This stage attracts significant research efforts, driven by the growing interest in video analytics and live streaming. \textcolor{black}{Ingestion-stage designs aim to improve analytics accuracy, encoding complexity, video quality, bandwidth utilization, and upload latency, which is especially important for interactive applications.} 

The \textit{\textsf{processing stage}}, which primarily involves internal operations within the media server, handles the storage and transformation of ingested video. Transformation tasks, such as video segmentation and transcoding, enable the pipeline to manage heterogeneity in network connectivity and device capabilities. Video segmentation divides the video into smaller chunks, while transcoding converts these chunks into multiple representations with different resolutions, bitrates, and frame rates. Numerous research efforts target the integration of transcoding with tasks at the ingestion and distribution stages to optimize pipeline performance.

\textcolor{black}{The \textit{\textsf{distribution stage}} handles video delivery from the media server to a user device, which decodes and plays back the content. In addition to ISPs providing network connectivity, this stage involves CDNs that disseminate video from their edge servers to user devices with low latency and high QoE. To address the heterogeneity of user devices and variable network conditions, a media player on the user device runs an ABR algorithm. Among the components of the end-to-end pipeline, ABR, QoE, and CDN aspects attract the most research efforts.}

\textcolor{black}{
\textit{\textsf{QoE models}} express users' subjective experience as a function of measurable \textit{\textsf{influence factors (IFs)}} like stall duration and video quality. They rely on subjective testing and user experience design. Network engineering ensures low latency and high bitrate, which QoE models account for directly or through other IFs. Data science enhances their predictive power with learning-based techniques.}

~
\vspace{-3mm}
\textcolor{black}{\subsection{2D streaming modes}\label{sec:video_streaming_modes}}

\textcolor{black}{
\textit{\textsf{Video on demand (VoD)}} is the most dominant of the two streaming modes considered in this survey. This mode closely aligns with real-world applications of major streaming platforms, such as Netflix, and specifically involves serving pre-stored video from a media server. The reliance on the media server effectively decouples the ingestion and distribution stages: while distribution operates in real time, ingestion occurs beforehand under less stringent latency constraints. As a result, VoD employs different communication designs at the ingestion and distribution stages.
}

\textcolor{black}{
\textit{\textsf{Live streaming}} refers to an increasingly popular variant that requires real-time operation of the entire pipeline from video capture to playback. "Real-time" is a relative notion where acceptable latency depends on the particular application. This survey considers HAS-based live streaming for applications such as live broadcasting. HAS improves its support for live streaming through a variety of techniques, such as reducing chunk duration, delivering a chunk in multiple fragments, and prefetching expected chunks by the CDN edge server.
}

~
\vspace{-4mm}
\subsection{\textcolor{black}{Streaming protocols}\label{sec:protocols}}

\textcolor{black}{
The current streaming ecosystem involves a large number of protocols with their popularity varying across the pipeline stages. 
Apple's \textit{\textsf{HTTP live streaming (HLS)}} constitutes the most popular protocol due to its dominance at the distribution stage. The main competitors of HLS at this stage are \textit{\textsf{dynamic adaptive streaming over HTTP (DASH)}}~\cite{Stockhammer2011}, which is an 
open standard maintained by the \textit{\textsf{moving picture experts group (MPEG)}}. The \textit{\textsf{real-time messaging protocol (RTMP)}}
maintains its prominence as the leading upload protocol at the ingestion stage. \textit{\textsf{Web real-time communication (WebRTC)}} is a newer protocol challenging the dominant role of RTMP at this stage. Whereas RTMP uses the \textit{\textsf{transmission control protocol (TCP)}} as its transport protocol, both WebRTC relies instead on the \textit{\textsf{user datagram protocol (UDP)}} to support low-latency upload. \cite{WOWZAmediasystems2019}~presents the usage of streaming protocols by 391 global broadcasters in sports, radio, gaming, and other industries. 
}

~
\vspace{-9mm}
\textcolor{black}{\subsection{Previous  surveys}\label{sec:previous_surveys}}

A large number of earlier surveys tackle the important topic of video streaming. Due to the complexity of the end-to-end streaming pipeline, these surveys often focus on individual stages or specific elements within a stage. For instance, \cite{bentaleb2019,kua2017,sani2017} concentrate on ABR algorithms at the distribution stage. \cite{barman2019,Juluri2016,zhao2017} address QoE in video streaming and emphasize QoE modeling, while \cite{Barakabitze2020}~deals with QoE management in novel network architectures. \cite{afzal2019}~surveys video streaming over multiple wireless paths. Whereas \cite{futureofCDN} discusses CDN support for video streaming and other traffic classes, \cite{Li2021}~covers cloud-based video streaming. In contrast to the previous surveys, our work offers a holistic overview of video streaming across the entire end-to-end pipeline. \textcolor{black}{In addition to CDN support, QoE, and ABR algorithms at the distribution stage, our survey also reports on advances in video streaming at the ingestion and processing stages.} Besides, we offer an up-to-date perspective by highlighting more recent research findings in the field.

~
\vspace{-3mm}
\textcolor{black}{\subsection{Related topics beyond the survey scope}\label{sec:beyond_scope}}

\textcolor{black}{
While this survey offers a new end-to-end pipeline perspective on HAS of long-form 2D videos over the best-effort networks, the rich area of video streaming contains related topics outside the survey scope. In particular, we do not report on P2P solutions exemplified by WebRTC~\cite{p2pwebrtc,Sredojev2015} where a camera-equipped device transmits video directly to a user device without any assistance from a media server. By deviating from the HAS pipeline and relying on UDP instead of TCP, such P2P solutions seek to provide ultra-low end-to-end latency for effective support of interactive applications, such as video conferencing~\cite{Jansen2018}. Because UDP does not provide CC, these P2P streaming systems implement their own CC algorithms. For instance, WebRTC employs \textit{\textsf{Google congestion control (GCC)}}~\cite{gcc} as its default CC algorithm. Although a P2P streaming system is able to port from TCP an existing CC algorithm, such as \textit{\textsf{CUBIC}}~\cite{cubic} or \textit{\textsf{bottleneck bandwidth and round-trip propagation time (BBR)}}~\cite{bbr}, these general-purpose CC algorithms do not cater specifically for the needs of video streaming. The work on streaming-specific CC algorithms includes \textit{\textsf{self-clocked rate adaptation for multimedia (SCReAM)}}~\cite{ccrfc} and \textit{\textsf{network-assisted dynamic adaptation (NADA)}}~\cite{ccnada}.}

\textcolor{black}{
Compared to long-form videos, streaming of short-form videos differs significantly in its requirements and solutions. With a common duration of 15 to 60~s, depending on the streaming platform, a short-form video requires much less storage and bandwidth, making it feasible to implement techniques such as prefetching the entire video~\cite{prefetch}, relying on progressive download instead of HAS~\cite{progdown}, \textcolor{black}{using equal-size rather than equal-duration chunks~\cite{netrav}, simplifying the ABR algorithm, or even transmitting the entire video at a single bitrate~\cite{onebit}. With a stronger emphasis on user engagement and interaction, short-form streaming designs explicitly account for user behavior, such as screen scrolling~\cite{duvas}. \textit{\textsf{Short-form video streaming and recommendation (SSR)}}~\cite{SSR} represents solutions that jointly optimize video recommendation and bitrate adaptation.} In short, short-form streaming is a vast topic deserving a separate survey.}

\textcolor{black}{
360-degree video streaming also faces distinct challenges. To deliver immersive experiences in \textit{\textsf{virtual reality (VR)}}, \textit{\textsf{augmented reality (AR)}}, and \textit{\textsf{mixed reality (MR)}}, 360-degree videos require specialized equipment for capture and playback. This includes camera arrays, omnidirectional cameras, curved screens, and \textit{\textsf{head-mounted displays~(HMDs)}}. The creation and presentation of seamless panoramic videos involve advanced stitching and projection methods~\cite{360complex}. To manage the higher storage, processing, and bandwidth requirements, 
360-degree video streaming employs tile-based~\cite{360tiles} and viewport-based~\cite{360viewport} techniques, which lie outside our survey scope.
}

\textcolor{black}{
Future Internet architectures, such as SDN and NDN, offer radically new opportunities for video streaming and other applications~\cite{fias}. Specifically, \textit{\textsf{in-transit computing}}~\cite{innetsurv} promises to make streaming more efficient by leveraging the processing capabilities of network devices along the delivery path. Our paper reviews video streaming designs within the current Internet architecture.
}
 
\vspace{-4mm}
\textcolor{black}{\section{Taxonomy}\label{sec:Methodologies}}
\vspace{-1mm}

\begin{figure}[t!]
    \centering
    \begin{subfigure}{\textwidth}
        \centering
        \includegraphics[trim=6.5cm 0cm 3.5cm 0cm,clip,width=\linewidth]{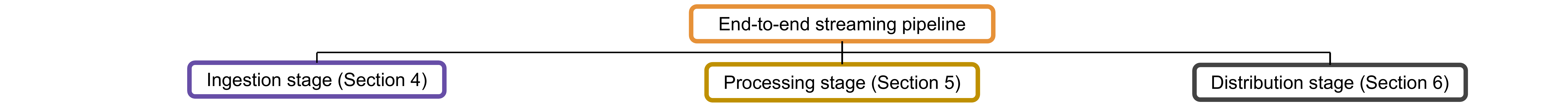}
        \vspace{-5mm}
        \caption{\textcolor{black}{top classification level based on pipeline stage}}
        \label{fig:3a}
        \vspace{-2mm}
    \end{subfigure}

    \vspace{1em}  

    \begin{subfigure}{0.47\textwidth}  
        \centering
        \includegraphics[trim=0cm 0cm 0cm 2.5cm,clip,width=\linewidth]{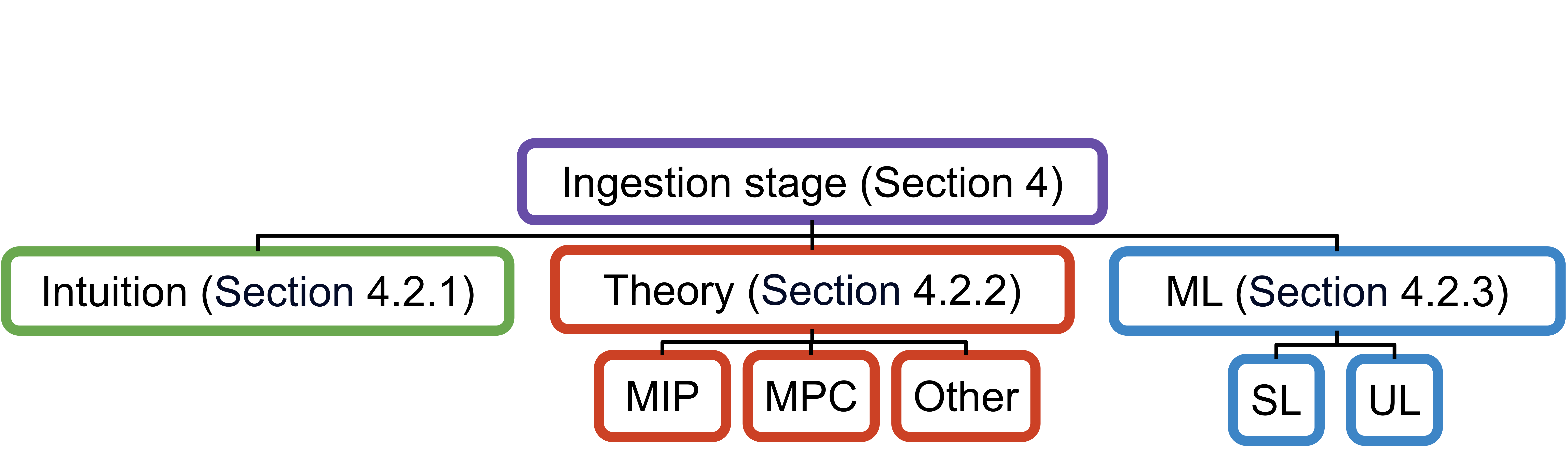}
        \vspace{-6mm}
        \caption{\textcolor{black}{methodology-based classification for the ingestion stage}}
        \label{fig:3b}
    \end{subfigure}
    \hfill
    \begin{subfigure}{0.47\textwidth}  
        \centering
        \includegraphics[trim=0cm 0cm 0cm 2.5cm,clip,width=\linewidth]{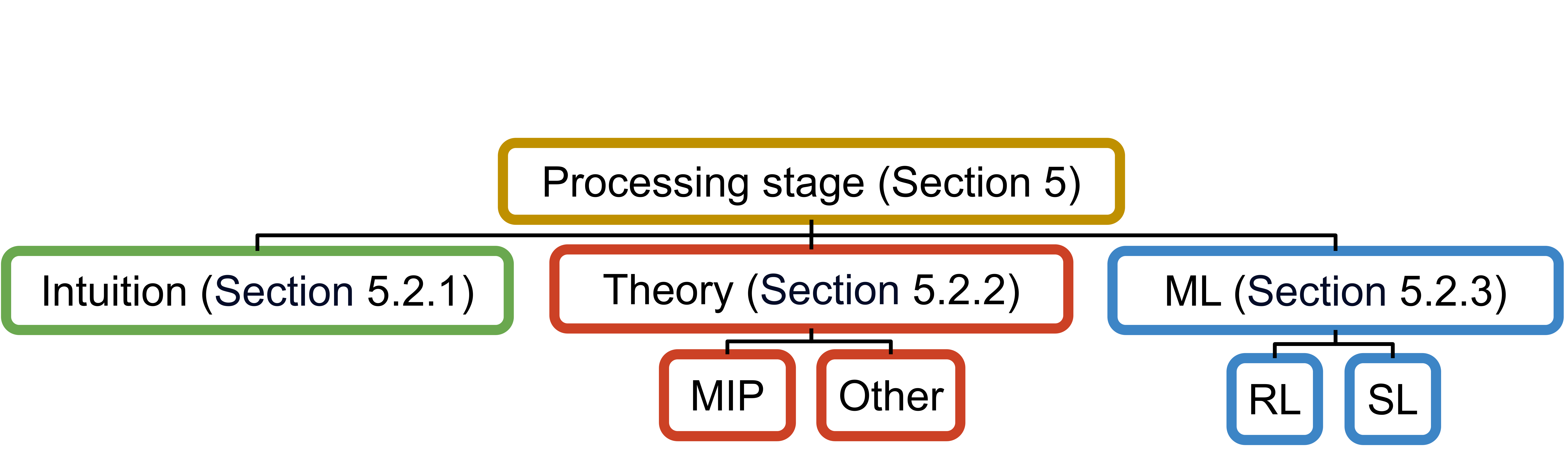}
        \vspace{-6mm}
        \caption{\textcolor{black}{methodology-based classification for the processing stage}}
        \label{fig:3c}
    \end{subfigure}

    \begin{subfigure}{\textwidth}
        \centering
        \vspace{1.5mm}
        \includegraphics[trim=0cm 1.5cm 0cm 0cm,clip,width=0.95\linewidth]{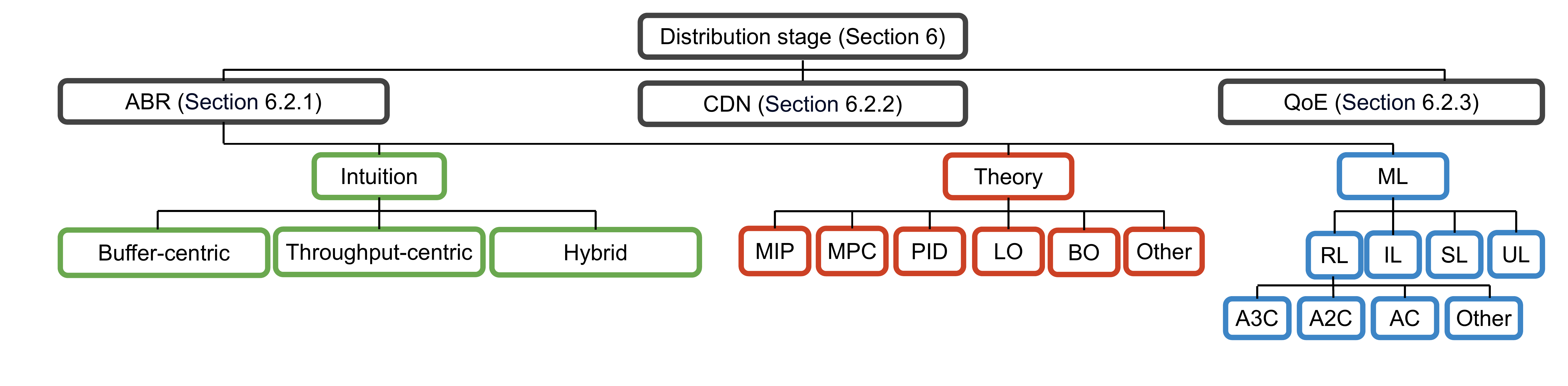}
        \vspace{-2mm}
        \caption{\textcolor{black}{methodology-based classification for ABR designs at the distribution stage of the end-to-end streaming pipeline}}
        \label{fig:3d}
    \end{subfigure}
    \vspace{-7mm}
    \caption{\textcolor{black}{Hierarchical taxonomy of the surveyed streaming designs.}}
    \label{fig: classification_scheme}
    \vspace{-5mm}
\end{figure}

\textcolor{black}{Figure~\ref{fig: classification_scheme} illustrates the taxonomy used in our survey. Figure~\ref{fig:3a} presents the top level of the hierarchy, which classifies streaming designs based on their operation at the {ingestion}, {processing}, or {distribution} \textit{\textsf{stage of the pipeline}}. The lower levels further classify designs based on their \textit{\textsf{solution methodology}}. Figures~\ref{fig:3b}, \ref{fig:3c}, and~\ref{fig:3d} provide these methodology-based taxonomies for the ingestion, processing, and distribution stages, respectively, with Sections~\ref{sec:Ingestion}, \ref{sec:Processing}, and~\ref{sec:Distribution} surveying the corresponding recent results. The classification scheme includes branches of varying breadth and depth to reflect the complexity and diversity of the reviewed works. For example, Figure~\ref{fig:3d} categorizes distribution-stage designs into {\em ABR}, {\em CDN}, and {\em QoE}  groups. Each final category lists works in chronological order and describes them based on \textit{\textsf{additional characteristics}}. While Section~\ref{sec:methodology-based_classification} elaborates on our methodology-based classifications, Section~\ref{sec:characteristics} discusses these additional characteristics.}

\vspace{-3mm}
\textcolor{black}{\subsection{Methodology-based classifications}\label{sec:methodology-based_classification}}

\textcolor{black}{Our methodology-based taxonomies differentiate designs according to their reliance on intuition, theory, or \textit{\textsf{machine learning (ML)}}, as discussed below.}

\subsubsection{\textcolor{black}{Intuition-based methods}\label{subsec:Intuition}}

In an intuition-based method, a human expert leverages domain knowledge and trial-and-error experimentation to develop a simple heuristic solution. It is common for an informal intuition-based method to undergo subsequent formal analysis, supplying insights into the underlying principles. An intuition-based heuristic might prove broadly applicable beyond the initial problem. A notable example is the \textit{\textsf{additive-increase multiplicative-decrease (AIMD)}} algorithm~\cite{AIMD}, originally designed for network CC and now employed widely in video streaming and other fields. 
\textcolor{black}{
For intuition-based ABR algorithms, we include a deeper level of classification that considers   \textit{\textsf{buffer-centric}}, \textit{\textsf{throughput-centric}}, and \textit{\textsf{hybrid}} categories, where ABR decisions rely on playback-buffer occupancy, network-bandwidth estimate, or both, respectively.}

~
\vspace{-8mm}
\subsubsection{\textcolor{black}{Theory-based methods}\label{subsec:Theory}}

A theory-based method abstracts specific details to formulate a problem within a general formal theory and systematically applies principles of rational logic to derive a solution, often with guarantees of correctness and performance. In comparison to intuition-based methods, the derived solution might be less intuitive or even counterintuitive. \textcolor{black}{\textit{\textsf{Mixed-integer programming (MIP)}} constitutes a prominent theory-based method for formulating and solving optimization problems~\cite{mip}. Control-theoretic techniques, such as \textit{\textsf{model predictive control (MPC)}}, \textit{\textsf{proportional-integral-derivative (PID)}} controllers~\cite{control_theory_book}, and \textit{\textsf{Lyapunov optimization (LO)}}~\cite{Lyapu}, commonly underpin solutions in video streaming. \textit{\textsf{Bayesian optimization (BO)}}~\cite{bo} represents a popular statistical optimization method. Our classification utilizes these MIP, MPC, PID, LO, and BO categories commensurately with the diversity of reviewed works: MIP and MPC at the ingestion stage, only MIP at the processing stage, and all five categories for ABR algorithms at the distribution stage. The sixth category, called \textit{\textsf{other}}, contains theory-based techniques applied less frequently in video streaming, such as \textit{\textsf{dynamic programming (DP)}}~\cite{bellmanoriginal}.
}
\vspace{-3mm}
\subsubsection{\textcolor{black}{ML-based methods}\label{subsec:ML}}

ML trains models on sample data to generalize and produce accurate predictions on new data, rather than following explicit instructions. The focus of
ML is on learning generalizable patterns and minimizing error on unseen samples, distinguishing it from theory-based methods that optimize solely for the given data. While promising better performance, ML raises concerns about higher overhead and poorer explainability. 
\textcolor{black}{We classify ML techniques into four categories based on their model training methodology
as \textit{\textsf{reinforcement learning (RL)}},  \textit{\textsf{imitation learning~(IL)}}, \textit{\textsf{supervised learning~(SL)}}, and \textit{\textsf{unsupervised learning~(UL)}}~\cite{sutton2018reinforcement}. At the distribution stage, our classification scheme further divides RL into \textit{\textsf{asynchronous advantage actor critic (A3C)}}, \textit{\textsf{advantage actor critic (A2C)}}, \textit{\textsf{actor critic (AC)}}~\cite{ac}, and \textit{\textsf{other}} methods. At the processing and ingestion stages, the reviewed works employ RL, SL, or~UL.
}

\vspace{-8mm}
\textcolor{black}{\subsection{Additional design characteristics}\label{sec:characteristics}}

\textcolor{black}{
In addition to applying the taxonomy from Figure~\ref{fig: classification_scheme}, we describe each reviewed design using a set of extra characteristics, which vary by design category.} {For every design, we specify its \textit{\textsf{core technique}}, a free-form characteristic of the design's main distinguishing trait, and \textit{\textsf{codec}} compatibility. For example, when discussing ML-based designs, the core technique characterizes the used model, such as a \textit{\textsf{decision tree~(DT)}}, \textit{\textsf{random forest~(RF)}}, \textit{\textsf{naive Bayes (NB)}}, \textit{\textsf{multilayer perceptron~(MLP)}}, \textit{\textsf{convolution neural network~(CNN)}}, \textit{\textsf{generative adversarial network~(GAN)}}, \textit{\textsf{autoencoder~(AE)}}, \textit{\textsf{generative pre-trained transformer~(GPT)}}, or another \textit{\textsf{deep neural network~(DNN)}}~\cite{MLbook,DLsurvey}.
}

\vspace{-3mm}
\textcolor{black}{\subsection{Main takeaways}}
\textcolor{black}{
The hierarchical taxonomy classifies streaming designs by pipeline stage (ingestion, processing, distribution), with the distribution stage additionally divided into ABR, CDN, and QoE categories, and by solution methodology (intuition, theory, ML). Intuition-based methods rely on heuristics and domain expertise, theory-based methods use formal logic with performance guarantees, and ML-based methods learn from data. Additional pipeline-wide and stage-specific characteristics provide further insights. With varying depth and breadth, this taxonomy captures the diversity and complexity of streaming research.
}

~
\vspace{-3mm}
\textcolor{black}{\section{Ingestion stage}\label{sec:Ingestion}}

~
\vspace{-5mm}
\textcolor{black}{\subsection{Background}}

\begin{figure}[t!]
\centerline{\includegraphics[trim=10 0 0 0,clip,width=\linewidth]{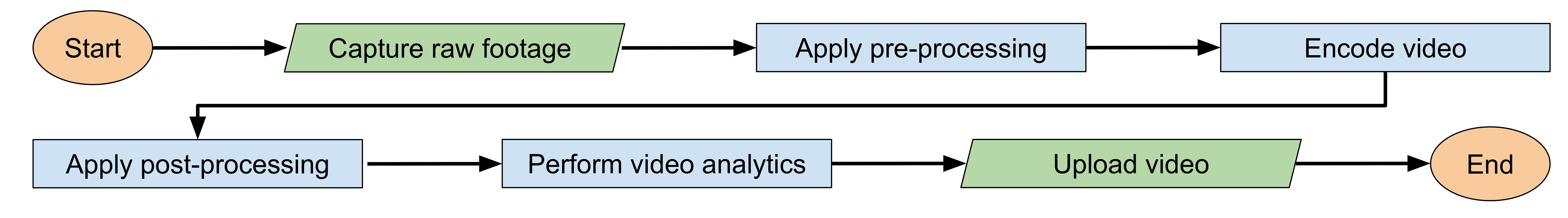}}
	\captionsetup{justification=centering}
 \vspace{-5mm}
	\caption{\textcolor{black}{The ingestion stage of the end-to-end VoD streaming pipeline.}}
	\label{fig:ingestion_flowchart}
 \vspace{-1mm}
\end{figure}

\textcolor{black}{
We proceed by providing additional background on ingestion,
with Figure~\ref{fig:ingestion_flowchart} illustrating 
this stage for VoD with a flowchart. The stage starts with the camera-equipped device capturing raw footage. Then, the device applies pre-processing, such as color correction and balancing, and encodes the video with a codec. After post-processing, such as artifact filtering, the ingestion stage provides optional support for video analytics, e.g., object detection and recognition. The stage concludes with uploading the encoded video to the media server. Hosting the media server in cloud infrastructure is increasingly common in both VoD and live streaming~\cite{cloudbook}.
}

\subsubsection{\textcolor{black}{Video encoding}\label{Sec:Ingestion_video compression}} 

Compression, which might occur during both ingestion and processing, is either lossy or lossless. Lossy compression reduces storage and bandwidth needs by discarding some information while maintaining high content quality. Spatial compression removes redundancy within a frame, e.g., by using \textit{\textsf{discrete cosine transform (DCT)}} and quantization, and encodes the result to reduce the bit count. Temporal compression, which is more computationally demanding, reduces redundancy across multiple frames through motion estimation and compensation. The codec or post-processing applies filtering to correct block boundaries, mosquito noise, ringing, and other artifacts introduced by lossy compression~\cite{video_compression_book}.

A compressed video involves different frame types. \textit{\textsf{Intra-frames (I-frames)}}, resulting from spatial compression, serve as reference points, prevent error accumulation, and facilitate video search.  \textit{\textsf{Predictive frames (P-frames)}}  use motion compensation based on previous frames, while \textit{\textsf{bipredictive frames (B-frames)}} leverage both preceding and following frames. A \textit{\textsf{group of pictures (GOP)}} refers to a sequence starting with an I-frame, followed by P-frames and B-frames. A single container format file stores the encoded video along with audio, synchronization, subtitle, and other metadata.

\textcolor{black}{
Video encoding is computationally intensive, with innovations aimed primarily at faster processing. Alongside algorithmic advances, hardware-accelerated encoding becomes more prevalent and offloads tasks to specialized components. Examples include \textit{\textsf{Nvidia encoder (NVENC)}}, which shifts video encoding from the central processing unit to the graphics processing unit (GPU), and \textit{\textsf{video coding engine (VCE)}}, a GPU-integrated unit dedicated to video compression.
}


\textcolor{black}{{\em Encoding parameters:}
A codec balances trade-offs between video quality, latency, and other performance metrics through various encoding parameters. Higher resolutions enhance image sharpness but demand more storage and bandwidth. For optimal results, the video resolution should match the display resolution. Frame rates typically range from 24 to 60~fps. Some applications, such as gaming, may require up to 120~fps~\cite{fpsstudy}. The GOP structure defines the number of frames per GOP and the spacing between keyframes (I-frames and P-frames). Larger GOPs with more B-frames reduce video size but increase processing complexity and latency.}

\textcolor{black}{{\em Prominent codecs:} \textit{\textsf{Advanced video coding (AVC)}} or
\textit{\textsf{H.264}} is one of the most widely used compression standards~\cite{H264}. It employs macroblocks and motion compensation for efficient video encoding. Key features include integer DCT, variable block-size segmentation, multi-frame inter-frame prediction, and in-loop deblocking filtering. H.264 remains the most popular codec due to its broad support across commercial devices~\cite{bitmovin2020}.}

\textit{\textsf{High efficiency video coding (HEVC)}} or \textit{\textsf{H.265}}, a successor to H.264, achieves up to 50\% better compression efficiency while maintaining the same video quality. \textcolor{black}{It replaces 16×16 macroblocks with \textit{\textsf{coding tree units (CTUs)}} up to 64×64 in size and uses both integer DCT and \textit{\textsf{discrete sine transform (DST)}}. HEVC simplifies deblocking filtering, making it easier to parallelize~\cite{hevc}.} Despite superior performance, adoption is slow due to royalty issues and limited browser support. 

\textit{\textsf{VP9}}, an open royalty-free codec developed by Google and utilized on YouTube\textcolor{black}{, employs 64×64 superblocks with \textit{\textsf{quadtree (QT)}} partitioning and intra-frame prediction with six oblique directions for linear extrapolation of pixels.} While less efficient in compression than H.265~\cite{confront1}, VP9 reduces encoding latency and enjoys broad browser support.

\textit{\textsf{AOMedia video 1 (AV1)}},  a royalty-free successor to VP9, diversifies coding options for better video input handling \textcolor{black}{and uses rectangular DCTs, asymmetric DSTs, and superblocks up to 128×128. It also employs in-loop and loop-restoration filters.} While AV1 incurs higher computational complexity to improve compression efficiency over H.265~\cite{av1over}, subjective tests of video quality show minimal differences~\cite{av1vshevcsubj}. 

\textit{\textsf{Scalable video coding (SVC)}} extends H.264 by enabling layered encoding into multiple streams, where enhancement layers build upon the base layer. These layers improve the frame rate, resolution, bitrate, or combinations thereof~\cite{svcgeneric}. 
Although less efficient in compression than H.264, SVC better manages highly variable bandwidth~\cite{svcperformance}. 

\textit{\textsf{Versatile video coding (VVC)}} or \textit{\textsf{H.266}}~\cite{Bross2021}, adopted in 2020, is a successor to H.265 that supports lossless and subjectively lossless compression. It aims to support a wide range of video applications through layered coding and flexible bitstream handling. VVC offers significant improvements in compression efficiency over H.265 and requires more computational resources~\cite{av1vshevc}. The royalty situation for VVC is still uncertain.

\textit{\textsf{Essential video coding (EVC)}}~\cite{evc}, introduced in 2020 as well, features innovations such as a binary-ternary tree structure, split unit coding order, and adaptive loop filter.  It improves compression efficiency by about 30\% over H.265, albeit with five times the computational complexity ~\cite{evcvvchevcav1}. EVC is available in both royalty-based and royalty-free profiles.

\textit{\textsf{Low complexity enhancement video coding (LCEVC)}}~\cite{lcevc} constitutes a novel approach to video enhancement. LCEVC adds an enhancement layer to a base layer encoded with a different codec, with an objective to reduce both encoding and decoding complexity.
\vspace{-4mm}
\subsubsection{\textcolor{black}{Perceptual compression}}
Unlike codecs that reduce statistical redundancy, perceptual video compression leverages properties of the human visual system to reduce the video size without compromising the perceived quality. It identifies 
\textit{\textsf{regions of interest (ROIs)}}
as spatial, temporal, or spatio-temporal areas critical for perception and  encodes ROIs losslessly, while applying stronger compression to less critical parts. \textcolor{black}{This process involves two phases: detecting ROIs with techniques ranging from user input to non-visual information~\cite{roi_survey}, and ROI-aware encoding, which might take place during pre-processing or actual encoding~\cite{roirateco2}.} 
\vspace{-4mm}
\subsubsection{\textcolor{black}{Super resolution (SR)}}\label{sec:super_resolution_background}
SR~\cite{surv_book} refers to a computer-vision task that reconstructs \textit{\textsf{high-resolution (HR)}} images from \textit{\textsf{low-resolution (LR)}} versions. In video streaming, SR reduces network-bandwidth consumption by transmitting LR frames and reconstructing HR video at the recipient.
While traditional SR relies on spatial-frequency substitution and geometric techniques, modern ML-based approaches employ GANs, CNNs, and other DNNs~\cite{surv_SR1}. \textcolor{black}{Despite improving video quality and bandwidth efficiency, ML-based SR techniques face challenges such as poor generalization, high parameter dimensionality, and balancing inference accuracy with speed.}

~
\vspace{-6mm}
\textcolor{black}{\subsection{Recent results}}

\textcolor{black}{
Recent works at the ingestion stage commonly tackle tasks such as video encoding, analytics, and upload. These studies evaluate the effectiveness of their solutions within the stage via metrics of bandwidth utilization, encoding complexity, video quality, analytics accuracy, upload latency,  and computational overhead. Additionally, some studies assess user experience by means of QoE models.
To achieve their goals, the reviewed designs explore various approaches, including assistance from SR, transport-layer signals, and edge servers.
}

\textcolor{black}{
This section, along with Table~\ref{tab:ing_table},  organizes our discussion of the recent works according to the methodology-based classifications presented in Section~\ref{sec:methodology-based_classification}: intuition, theory (MIP, MPC, and other), and ML (SL and UL). In addition to the \textit{\textsf{core technique}} and \textit{\textsf{codec}} characteristics explained in Section~\ref{sec:characteristics}, Table~\ref{tab:ing_table} describes each ingestion-stage design based on five stage-specific binary characteristics: (1)~\textit{\textsf{SR}} usage, (2)~ utilization of a well-defined \textit{\textsf{QoE model}} in design or evaluation, (3)~reliance on \textit{\textsf{transport-layer signals}}, (4)~leverage of \textit{\textsf{edge infrastructure}}, and (5)~\textit{\textsf{bandwidth-efficiency evaluation}} in the reviewed work.
}

\input{1_Ingestion_table}

\subsubsection{\textcolor{black}{Intuition-based methods}}

\textcolor{black}{Guided by measurements of TCP uplink throughput in a radio access network,  \cite{Lottermann2015}~intuitively reduces the number of bitrate levels in the encoding ladder and thereby conserves bandwidth.} This technique combines real-time and historical throughput data, using the former for ongoing sessions and the latter at the start of sessions or during handovers. \cite{Wilk2016}~proposes dynamic selection of the upload protocol by a mobile broadcasting application. The application considers latency, join-time, goodput, and overhead metrics, picks one of them, evaluates this metric in real time, and periodically decides whether to switch to another upload protocol. While this method performs as well as the best protocol for each individual  metric, the switching between protocols incurs undesirable delay. \cite{Park2017}~monitors the average inter-arrival time of video frames and dynamically adjusts the encoding rate on a camera-equipped mobile device via the AIMD algorithm. By increasing the average encoding rate and decreasing the packet loss, the algorithm improves real-time upstreaming under changing network conditions. \textit{\textsf{NeuroScaler}}~\cite{neuroscaler} enhances the scalability of SR-based live streaming by lowering both overhead and encoding time of SR. The design includes a novel scheduler and enhancer of the anchor frames used by SR. The anchor scheduler leverages codec-level information to select the anchor frames in real time without any neural inference. The anchor enhancer complements a video codec with a simple image codec and employs the latter for compression of the anchor frames only. 
\vspace{-2mm}
\subsubsection{\textcolor{black}{Theory-based methods}}

 \textcolor{black}{\textit{\textsf{Vantage}}~\cite{Ray2019} refers to a MIP-based approach that targets social live streams and improves QoE for time-shifted viewers through frame retransmissions.} When bandwidth allows, it retransmits earlier frames at a higher bitrate, enhancing the experience for viewers watching with time shifts. Vantage employs MIP for retransmission scheduling.
 \textcolor{black}{\textit{\textsf{LiveSRVC}}~\cite{chen2021} is an MPC-based solution for live-stream ingestion, aiming to decrease bandwidth usage and latency via SR.} It compresses I-frames at the camera side and trains an SR model online to reconstruct them on the server. Guided by estimated uplink bandwidth, SR processing time, and accuracy, LiveSRVC uses MPC to select the I-frame compression ratio and chunk bitrates. 
 
 \textcolor{black}{Other theory-based ingestion works include \cite{Siekkinen2017}, which, similar to Vantage, strives to maximize video quality in live streaming for multiple clients with heterogeneous upload latencies.} The design involves a series of algorithms that leverage a greedy low-complexity DP-based approach. Conversely, the \textit{\textsf{content harvest network (CHN)}}~\cite{Pang2019} achieves both low latency and efficient bandwidth utilization during ingestion by employing edge devices as relays to direct traffic from broadcasters to servers. To determine the path for each broadcaster, CHN employs two strategies on different time scales. Whereas finding a globally optimal path
 is a \textit{\textsf{nondeterministic polynomial time (NP)}} and NP-hard problem, a centralized server periodically solves it via a polynomial-time greedy rounding algorithm. \cite{8812206}~selects both the upload server and encoding bitrate to jointly maximize the video rate and minimize end-to-end latency. It develops algorithms for both one-hop-overlay and full-overlay architectures. The one-hop-overlay algorithm is an optimal polynomial-time solution. The paper proves NP-completeness of the full-overlay problem and solves it with an efficient heuristic solution based on convex relaxation.

\textcolor{black}{\textit{\textsf{DNN-driven streaming (DDS)}}~\cite{du2020} refers to a theory-based solution where the camera-equipped device optimizes bandwidth usage across two streams to enhance inference accuracy while minimizing bandwidth consumption in analytics applications.} The first stream transmits low-quality video to the server, which identifies ROIs for DNN inference. The second stream provides high-quality video for the detected ROIs, improving inference accuracy while managing bandwidth efficiently. DDS applies a Kalman filter to estimate base bandwidth and adjusts bandwidth usage by tuning the resolution and \textit{\textsf{quantization parameter~(QP)}}. With a similar focus on camera uploads,  \textit{\textsf{LiveNAS}}~\cite{kim2020} employs SR for high-quality live streaming. Along with the live video, the camera-equipped device also uploads high-quality frame patches. The server utilizes these patches to train a DNN for SR in real time. LiveNAS allocates upload bandwidth between the live video and patches by means of gradient ascent to maximize both video quality and DNN accuracy, while minimizing overhead for ingest clients.

~
\vspace{-8mm}
\subsubsection{\textcolor{black}{ML-based methods}}

\textcolor{black}{\cite{saliencyandML,cai2020}~present SL-based codecs for perceptual compression, targeting improvements in coding efficiency.} Compared to standard codecs, these designs increase video quality and decrease storage requirements while decreasing the encoding speed. \cite{saliencyandML}~extends the H.265 codec by incorporating a hybrid compression algorithm that employs a CNN for spatial saliency and then extracts temporal saliency from motion information in the compressed domain. \cite{cai2020}~introduces an ROI codec that combines CNNs with an entropy codec to achieve better encoding efficiency than previous ROI codecs, though its decoding performance is less effective.
\textit{\textsf{CrowdSR}}~\cite{luo2021} enhances live streaming from low-end devices via SR-based video uploading. It periodically trains an SR model with high-quality video patches from similar content broadcasters. CrowdSR outperforms existing counterparts in regard to the \textit{\textsf{peak signal-to-noise ratio (PSNR)}}~\cite{psnr} and \textit{\textsf{structural similarity index measure (SSIM)}}~\cite{ssimcite}. In contrast, \textit{\textsf{DIVA}}~\cite{Xu2021} improves video analytics efficiency by leveraging both camera-equipped device and server. It processes only key video frames on the camera-equipped device to avoid unnecessary uploads. Utilizing the sparse analytical data, the server trains CNNs, specifically variants of AlexNet~\cite{alexnet}, and sends them back to the camera-equipped device to identify I-frames for upload. This iterative approach enhances analytics performance and operates 100 times faster than real-time video.   \textcolor{black}{Recent work on neural codecs includes \textit{\textsf{MobileCodec}}~\cite{interframe}, which adopts a DNN architecture and SL to support efficient coding for mobile devices. It features an inter-frame module with fully convolutional operations, asymmetrical design for faster real-time decoding, and activation quantization with simulated straight-through gradient estimation.}

\textcolor{black}{Applying UL to perceptual compression, \textit{\textsf{DeepFovea}}~\cite{kaplanyan2019} proposes foveated coding that strengthens compression for areas outside the fovea.} The codec employs a GAN to reconstruct realistic peripheral video from a minimal set of frame pixels. It operates quickly enough for HMDs and delivers superior perceptual quality in subjective evaluations. In contrast, \textit{\textsf{Reducto}}~\cite{li2020a} aims to reduce bandwidth consumption in UL-based video analytics. It tracks basic features, such as pixel and edge differences, and identifies relevant features for specific queries. Reducto relies on $k$-means clustering to establish a dynamic threshold for frame filtering at the camera-equipped device. By filtering out less important frames, it reduces upload traffic while maintaining analytics accuracy.
\textcolor{black}{
Among the latest advancements in neural codecs,  \cite{datascalable}~introduces a data-scalable codec that employs AEs trained with UL and custom loss function. This codec enhances compression quality with each new packet received and achieves the highest quality when there is no packet loss.
}

~
\vspace{-7mm}
\subsection{Main takeaways}

The review of recent research on ingestion-stage designs reveals a strong focus on live streaming. This emphasis stems from the growing importance and significant technical challenges of the live mode. The stricter end-to-end latency constraints of live streaming affect the ingestion stage and promote a trend toward integrated end-to-end streaming solutions. Another trend is the increasing computational role of camera-equipped devices able to offload processing from media servers. This offloading delivers faster video analytics and decreases upload bandwidth consumption. ML-based methods are increasingly prominent at the ingestion stage, either as core algorithms or supporting components. In particular, ML-based SR methods receive considerable attention and success, with a prevalence of adapting existing models and effective training strategies rather than developing new ML techniques.

\begin{figure}[t!]
\centerline{\includegraphics[trim=10 70 150 0,clip,width=\linewidth]{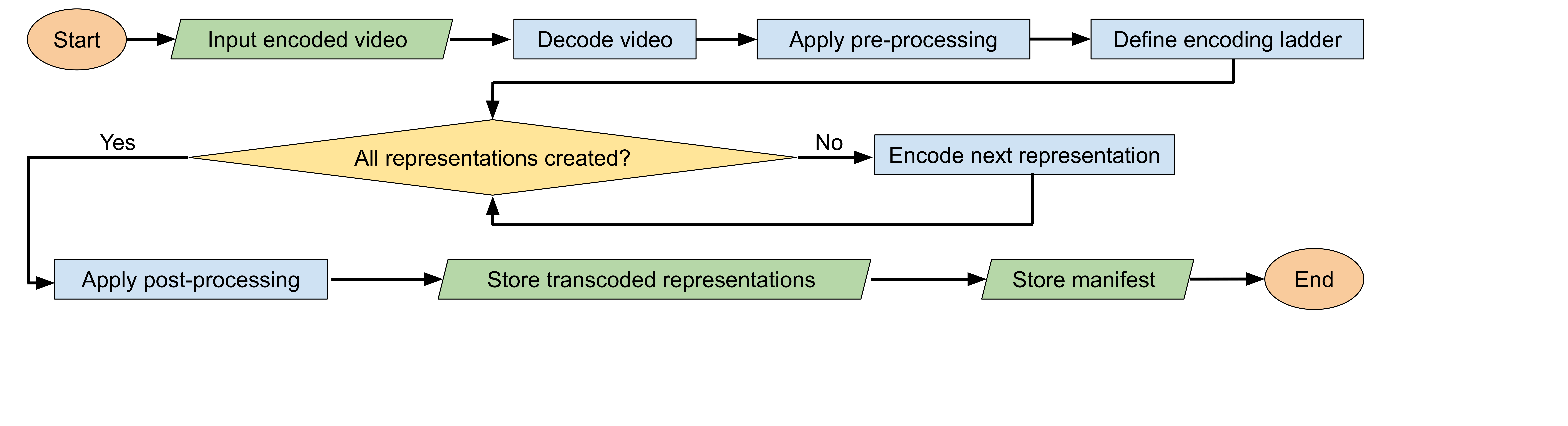}}
	\captionsetup{justification=centering}
 \vspace{-5mm}
	\caption{\textcolor{black}{Transcoding at the processing stage of the end-to-end streaming pipeline.}}
	\label{fig:transcoding_flowchart}
 \vspace{-4mm}
\end{figure}

~
\vspace{-7mm}
\textcolor{black}{\section{Processing}\label{sec:Processing}}

~
\vspace{-5mm}
\textcolor{black}{\subsection{Background}}

The processing stage lies between ingestion and distribution in the end-to-end streaming pipeline. It operates on dedicated or cloud servers and performs various tasks to support the adjacent stages. The essential task at the processing stage of the HAS pipeline is \textit{\textsf{transcoding}}~\cite{transcoding1}, which converts encoded video into multiple representations. Since transcoding produces compressed videos, it shares similarities with the video compression performed during the ingestion stage, making the background information in Section~\ref{Sec:Ingestion_video compression} relevant. However, there are key differences. While the primary goal of encoding on the camera-equipped device is to efficiently utilize the ingestion bandwidth, transcoding leverages the superior  computational and storage resources of the media server to create compressed videos suitable for distribution to a wide range of user devices. 

\textcolor{black}{
Figure~\ref{fig:transcoding_flowchart} presents a flowchart of transcoding. After receiving encoded video as input, the task decodes the video and applies pre-processing, such as noise filtering. Then, the task defines an encoding ladder
by specifying the target bitrate, resolution, and frame rate of each representation. Transrating and transsizing refer to transcoding where the generated representations differ only in their bitrate or resolution, respectively. For each rung of the encoding ladder, the process re-encodes the decoded video to create a corresponding new representation. After post-processing, such as subtitle embedding, the transcoding task stores the created representations on the media server and records the encoding ladder in a manifest file. 
}

Alongside transcoding, which is intrinsic to the HAS pipeline and directly impacts end-to-end streaming performance, the processing stage performs a variety of auxiliary tasks. {Video splitting} divides the video into smaller chunks for HTTP compatibility, typically ranging from 2 to 10~s in duration, with this variation significantly affecting the quality of video streaming~\cite{cs1}. {Video editing} alters the video content, e.g., by adding advertisements or removing censored material. Traditionally carried out at the processing stage, {video analytics} employs techniques from computer vision for object detection and image segmentation, classification, and recognition. {Video storage} on the media server is particularly important for VoD, where videos need to remain available over extended periods.

\textcolor{black}{\subsection{Recent results}}

\textcolor{black}{
Our review of recent research at the processing stage focuses on its main task of transcoding. These studies typically aim to reduce processing time, energy consumption, storage needs, and bandwidth usage. 
}

\textcolor{black}{
Again, we present the reviewed works according to the methodology-based classifications outlined in Section~\ref{sec:methodology-based_classification}: intuition, theory (MIP and other), and ML (RL and SL). Besides the \textit{\textsf{core technique}} and \textit{\textsf{codec}}, which are relevant across all stages, Table~\ref{tab:proc_table} also characterizes the processing-stage designs based on: (1)~optimization \textit{\textsf{type}} as online, offline, or hybrid, (2)~processing, energy, storage, or bandwidth as \textit{\textsf{performance}} improvement objectives, and (3)~explicit consideration of edge or CDN \textit{\textsf{infrastructure}}.
}

\subsubsection{\textcolor{black}{Intuition-based methods}} 

\textcolor{black}{To support fast low-complexity transcoding from H.264 to H.265, \cite{f264to265}~employs intuitive statistics-driven heuristics for different types of  \textit{\textsf{coding units (CUs)}}.} These heuristics allow for early termination of CU partitioning and prediction unit mode selection. \cite{criptotra}~deals with transcoding video streams encrypted in the H.264 or H.265 formats. Because decrypting and re-encrypting these streams introduces significant latency, this work 
develops a joint crypto-transcoding scheme that enables transcoding of encrypted video streams without decrypting them or exposing the decryption key at intermediate devices.  To reduce energy consumption on mobile devices, \textit{\textsf{environment-aware video streaming optimization (EVSO)}}~\cite{evso} 
considers the device's battery status and generates encoding ladders that adjust the frame rate of different video chunks based on a new metric of perceptual similarity.

\input{2_Processing_table}

\vspace{-3mm}
\subsubsection{\textcolor{black}{Theory-based methods}} 

\textcolor{black}{To minimize storage and processing requirements, both \textit{\textsf{light-weight transcoding at the edge (LwTE)}}~\cite{lwte} and \textit{\textsf{adaptive bitrate ladder optimization for live video streaming (ARTEMIS)}}~\cite{ARTEMIS} rely on MIP.}
In LwTE, the edge server performs partial transcoding based on the optimal CU partitioning structure received from the origin server. By applying binary search to a \textit{\textsf{mixed-integer linear programming (MILP)}} formulation, LwTE heuristically distinguishes between popular and unpopular video chunks. For unpopular chunks, it stores only the highest bitrate level and generates lower bitrate levels on the fly through metadata-accelerated transcoding. In contrast, ARTEMIS dynamically defines the encoding ladder for \textcolor{black}{a live streaming session} by considering content complexity, network conditions, and detailed client feedback in a standard format~\cite{cmcd}. ARTEMIS advertises many representations via a mega-manifest file and employs MILP to select a smaller subset of these representations for the encoding ladder. \textcolor{black}{Similarly, \textit{\textsf{adaptive bitrate ladder optimization for multi-live HAS (ALPHAS)}}~\cite{alphas} constructs encoding ladders for multiple live streams by dynamically generating a content-aware bitrate ladder for each stream. It accounts for encoder computational capabilities, CDN bandwidth constraints, and stream prioritization, integrates the mega-manifest concept with real-time \textit{\textsf{video multimethod assessment fusion (VMAF)}}~\cite{vmafcite} prediction, and solves an \textit{\textsf{integer linear programming (ILP)}} formulation by leveraging its submodular properties.}

\textcolor{black}{Other theory-based works include~\cite{cdntranscoding}, where a CDN performs online just-in-time transcoding of a video chunk to the needed bitrate only when a user requests it.} This design relies on a Markov model to predict the bitrate requested for the next chunk, enabling the CDN to start delivering the transcoded chunk immediately upon receiving the request.
\cite{nettrans}~explores context-aware encoding and formulates encoding-ladder definition as an optimization problem that models the client's bandwidth estimates and viewport sizes as stationary random processes. 
To support energy-efficient transcoding, \cite{transenergy}~selects between three options: offline transcoding, online transcoding, and serving the chunk at a lower than requested bitrate. The selection seeks to maximize video quality within a limit imposed on the total transcoding time, formulates a knapsack-like problem, and solves the problem via a greedy heuristic.

\subsubsection{\textcolor{black}{ML-based methods}}

\textcolor{black}{\textit{\textsf{MAMUT}}~\cite{mamut} and  \textit{\textsf{DeepLadder}}~\cite{Rl_ladder} are RL-based designs for efficient real-time transcoding.} MAMUT employs multi-agent \textit{\textsf{Q-learning (QL)}} in an environment with multiple users, where three agents collaboratively adjust the number of encoding threads, QP, and processor frequency. This optimization seeks to maximize a reward function that combines the frame rate, bitrate, PSNR, and power consumption. On the other hand, DeepLadder leverages content features, available bandwidth, and storage costs to transcode each chunk according to an encoding ladder defined via a dual-clipped version of
\textit{\textsf{proximal policy optimization (PPO)}}.

\textcolor{black}{\cite{transratingpap,rftrans}~apply SL to limit the encoder's parameter search and thereby reduce transcoding time.} \cite{transratingpap}~employs DTs to constrain the maximum CTU depth, aiming to balance transcoding time, energy consumption, and video quality. \cite{rftrans}~accelerates cascaded pixel-domain transcoding by employing two RF classifiers to set upper and lower limits on the CTU depth. 
\textit{\textsf{Fast video transcoding time prediction and scheduling (FastTTPS)}}~\cite{fastttps} considers features of source videos, trains an MLP to predict transcoding time, and leverages the predictions to schedule parallel executions of transcoding tasks.
 \textit{\textsf{HEVC-based quadtree splitting (HEQUS)}}~\cite{nbtrans} reduces the encoder's parameter search for transcoding from H.265 to VVC. It trains NB classifiers to partition the first QT level into 128×128 blocks and uses the H.265 CU partitioning to guide QT splitting decisions for 64×64 blocks and lower levels.

~
\vspace{-7mm}
\subsection{Main takeaways}

Recent research efforts at the processing stage put a key focus on faster processing and lower power consumption. In particular, transcoding acceleration enables on-the-fly definition of encoding ladders, which not only decreases storage demands but also aligns with the growing trend toward live streaming. The explicit consideration of distribution-stage infrastructure, such as CDN or edge servers, reflects a closer integration across pipeline stages. ML-based methods are increasingly prominent in processing-stage designs and, in contrast to the ingestion stage, tend to employ simple models rather than deep networks. Additionally, most designs assume the use of H.264 or H.265 codecs rather than more advanced options.

~
\vspace{-7mm}
\textcolor{black}{\section{Distribution}\label{sec:Distribution}}

\begin{figure}[t!]
	\centerline{\includegraphics[trim=4.8cm 10.3cm 5cm 2.5cm,clip,width=\linewidth]{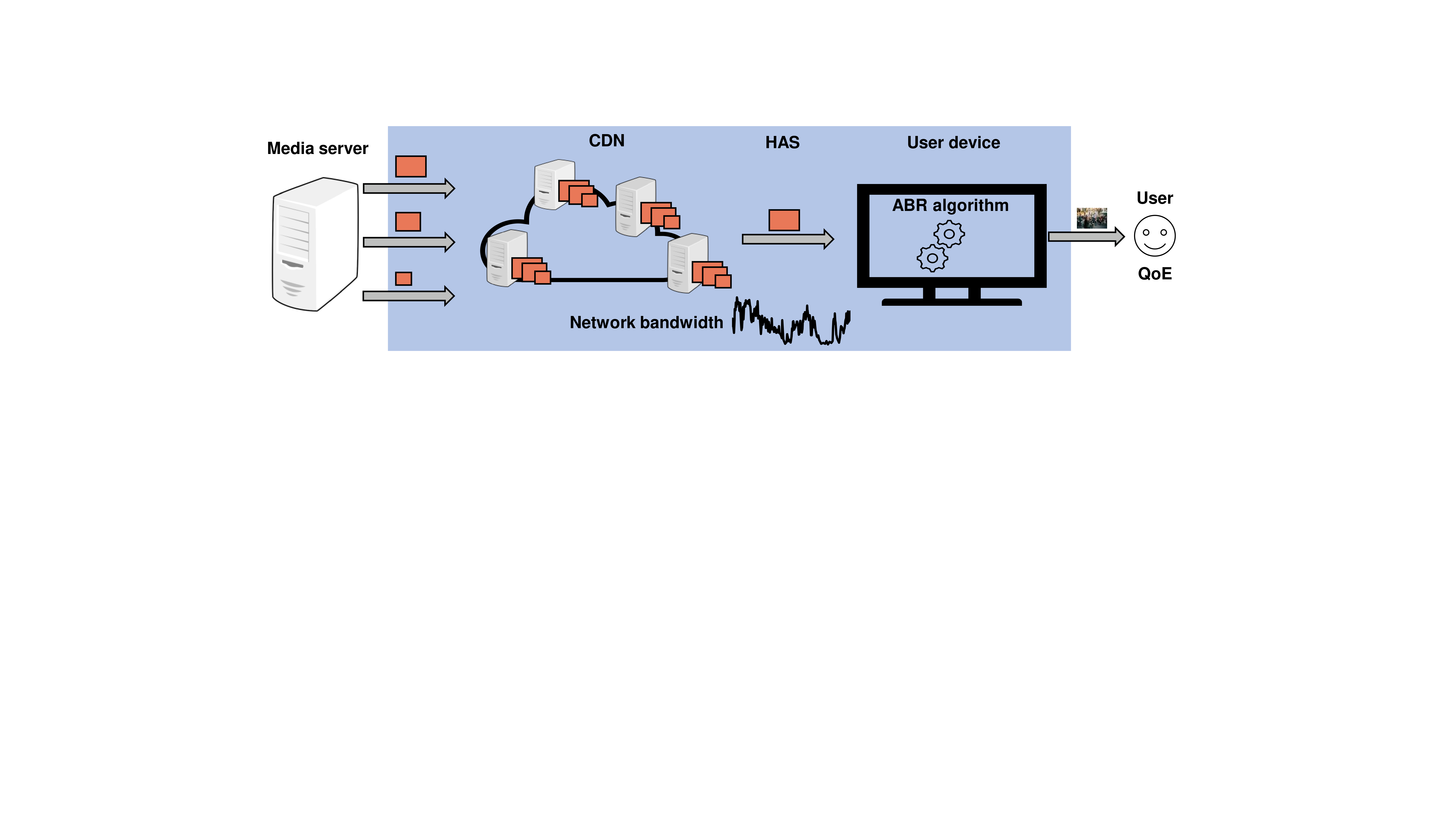}}
	\captionsetup{justification=centering}
 \vspace{-6mm}
	\caption{\textcolor{black}{The distribution stage of the end-to-end VoD streaming pipeline.}}
	\label{fig: distribution_picture}
 \vspace{-2mm}
\end{figure}

~
\vspace{-5mm}
\textcolor{black}{\subsection{Background}}

The end-to-end streaming pipeline concludes with the distribution stage, which delivers the requested video to the user device and plays it on the screen. \textcolor{black}{Figure \ref{fig: distribution_picture} illustrates the distribution stage of HAS for VoD.} At this stage, the user device requests one video chunk at a time from the CDN, which caches each chunk in multiple representations provided by the media server. The CDN supports scalable low-latency delivery by utilizing its extensive network of edge servers spread across different geographical regions. The ABR algorithm on the user device dynamically chooses the appropriate representation for the next requested chunk based on predictions of varying network bandwidth. This algorithm aims to balance uninterrupted playback with high video quality, ultimately ensuring high QoE for the user. Live streaming employs shorter chunks, downloads them from the camera-equipped device to the user device in real time, \textcolor{black}{and imposes more stringent requirements on distribution, prompting different approaches to CDN support and QoE improvement. This survey focuses on the key ABR, CDN, and QoE aspects of the distribution stage.}

\subsubsection{\textcolor{black}{ABR algorithms}}

\textcolor{black}{The ABR algorithm dynamically selects the chunk representation and serves as a cornerstone of HAS, with HLS and DASH being the predominant HAS protocols. 
While HLS commonly employs a chunk duration of 6~s (10~s originally) and is compatible with the H.264 or H.265 codecs, DASH typically has a chunk duration between 2 and 10~s and is codec-agnostic. Our survey focuses on the prevailing HAS approach that uses client-side ABR algorithms.}

\begin{figure}[t!]
	\centerline{\includegraphics[trim=0.5cm 0cm 0cm 0cm,clip,width=\linewidth]{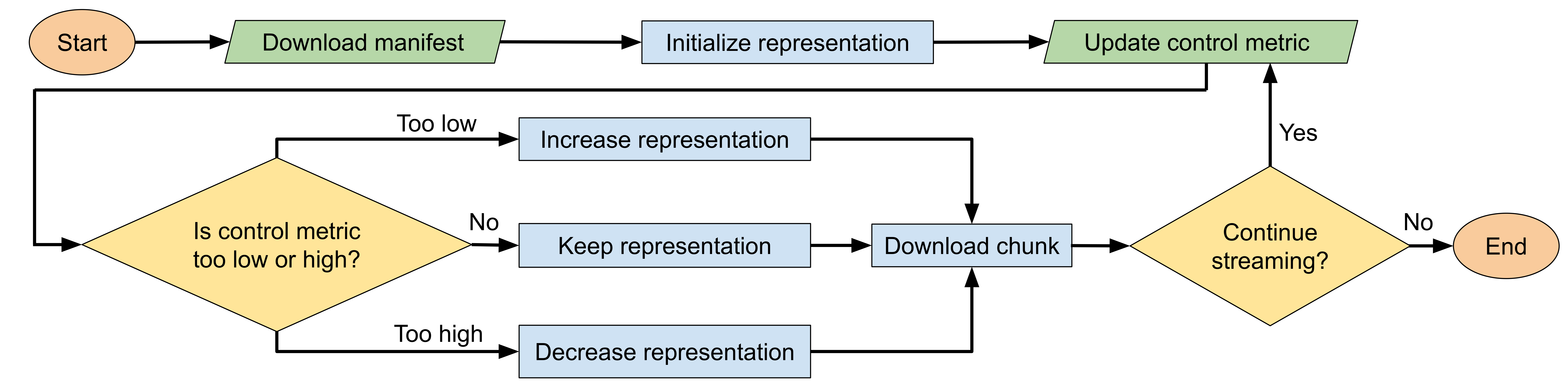}}
	\captionsetup{justification=centering}
 \vspace{-3mm}
	\caption{\textcolor{black}{The ABR algorithm.}}
	\label{fig:ABR_flowchart}
 \vspace{-4mm}
\end{figure}

\textcolor{black}{
Figure~\ref{fig:ABR_flowchart} depicts the ABR algorithm as a flowchart. At the start of the streaming session, the client downloads a manifest file from the media server.
The manifest includes an encoding ladder that describes the available representations for each video chunk in terms of their bitrate, resolution, and frame rate. The ABR algorithm updates a control metric based on the monitored network conditions. For example, the control metric is typically playback-buffer occupancy  and
network-bandwidth estimate in, respectively, buffer-centric and throughput-centric ABR algorithms. If the control metric indicates that the current representation is too low, the ABR algorithm increases the representation for the next chunk to enhance video quality. If the control metric is too high, the algorithm decreases the representation to avoid video stalling and rebuffering, which occur when chunks arrive too late for smooth playback. Otherwise, the representation remains unchanged. In all three cases, the client downloads the next chunk in the selected representation. This cycle of updating the control metric, selecting the appropriate representation, and downloading the chunk continues  until the end of the streaming session.
}

\textcolor{black}{Representation selection is challenging due to a priori unknown network conditions, mismatches between the manifest-file descriptions and actual chunk bitrates, large gaps between the bitrates of adjacent representations, and conflicting performance objectives. Because optimal ABR control is an NP-hard problem~\cite{hindsight}, practical ABR algorithms employ various heuristics, e.g., predicting the available network bandwidth from the client's historical throughput measurements.}

~
\vspace{-8.1mm}
\subsubsection{CDN support}

A CDN refers to a system of cache servers distributed across wide geographical areas to improve the performance of content delivery from CPs to end users~\cite{Pathan2006}. The CDN stores videos and other content collected from CPs' origin servers in cache servers placed near users, reducing network traffic and enabling low-latency content delivery~\cite{algorithmicnuggets}. Though originally optional, CDNs are indispensable in the modern Internet ecosystem and handle estimated 56\% and 72\% of all Internet traffic in 2017 and 2022, respectively~\cite{Cisco2019}.

Economic relationships with CPs form the basis for 
classifying CDNs as public, private, or hybrid. A public CDN, e.g. Akamai~\cite{akamai}, acts as a third party and charges the CPs for its content-delivery services. A private CDN belongs to the same organization as the CPs utilizing it, while a hybrid CDN serves both internal and external CPs. Due to CDNs' differences in scalability, pricing, and QoE across regions and time~\cite{cdncomparison}, CPs often deliver content over multiple CDNs.

Standards such as \textit{\textsf{common media client data (CMCD)}}~\cite{cmcd} and \textit{\textsf{common media server data (CMSD)}}~\cite{cmsd}, introduced in 2020 and 2022 respectively, enable information exchange between a CDN and clients to support data analysis and QoE monitoring. \textcolor{black}{Additionally, edge infrastructure extends the original CDN concept by involving network operators in content caching and offers new options for video streaming~\cite{genericedge}. }

~
\vspace{-7.5mm}
\subsubsection{\textcolor{black}{QoE}\label{subsec: qoe}}

\textcolor{black}{
\textcolor{black}{In contrast to the earlier notion of QoS, which encompasses individual network-level metrics such as packet loss, latency, and throughput,} QoE captures 
the user's subjective satisfaction with the overall performance of a streaming service~\cite{QoEdef}. 
QoE is crucial for streaming platforms because user satisfaction strongly correlates with customer attraction and retention and, ultimately, provider revenues. However, user perception of service performance is complex and depends on numerous IFs\textcolor{black}{, such as network bandwidth, latency, and video quality}~\cite{originalIFclass}.}

\textcolor{black}{
Assessing QoE is challenging due to its subjective interdisciplinary nature. Direct measurement typically involves subjective tests where users rate their streaming experience. These tests typically take place in controlled lab environments and follow well-established protocols informed by user experience design~\cite{zhao2017}. Online crowdsourcing improves testing scalability and weakens control over experimental settings~\cite{crowdbook}. The predominant approach to QoE evaluation is indirect and relies on subjective tests to build a QoE model that expresses QoE as a function of objectively measurable IFs. \textcolor{black}{Data science enhances the predictive power of QoE models via ML-based methods.} QoE models commonly represent QoE in terms of the \textit{\textsf{mean opinion score (MOS)}}~\cite{mos}, the average rating given by users in a subjective study.} 

\begin{figure}[t!]
\centerline{\includegraphics[trim=10 0 0 0,clip,width=\linewidth]{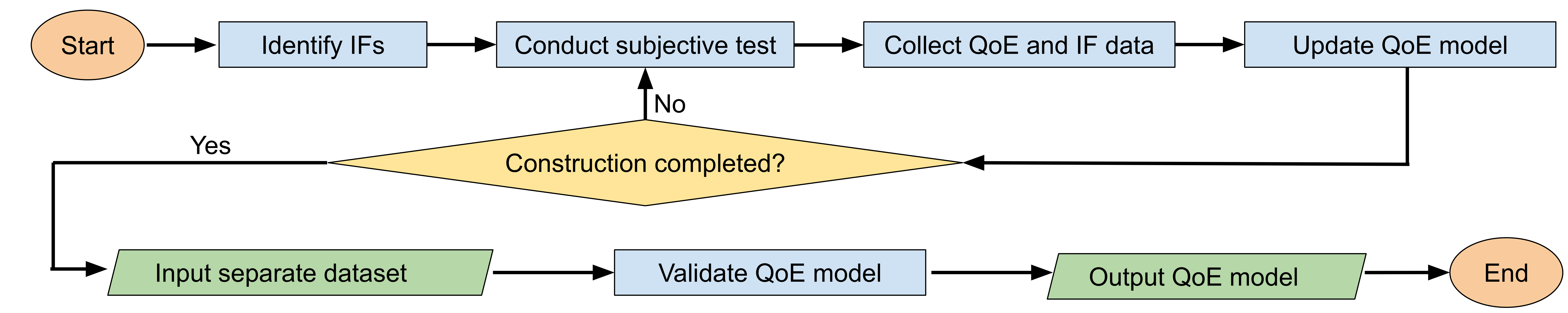}}
	\captionsetup{justification=centering}
 \vspace{-4mm}
	\caption{\textcolor{black}{QoE modeling.}}
	\label{fig:qoe_flowchart}
 \vspace{-3mm}
\end{figure}

\textcolor{black}{
Figure~\ref{fig:qoe_flowchart} illustrates \textit{\textsf{QoE modeling}} that constructs a QoE model iteratively. The process identifies the IFs of the QoE model and enters a cycle of conducting a subjective test, recording the respective IF values, collecting a user-provided QoE score, and mapping the IFs to the score to update the QoE model. This iterative refinement allows the current model to inform the configuration of the subsequent test, thereby reducing the number of subjective tests needed to develop an accurate QoE model~\cite{iQoE}. After the construction is complete, the process utilizes a separate dataset to validate the model and outputs the validated QoE model.
}

\textcolor{black}{
Once constructed, a QoE model supports automatic QoE computation based on the objectively measurable IFs, eliminating the need for human feedback and enabling QoE evaluation at scale. \textcolor{black}{Existing QoE models vary widely in terms of the IFs considered and the methods used for construction~\cite{originalmodelqoeclass}. Despite the significance of QoE and its models, their treatment often lacks standardization and rigor, creating opportunities for improvement~\cite{qoepitfalls}.}}

~
\vspace{0mm}
\textcolor{black}{\subsection{Recent results}}

~
\vspace{-6mm}
\subsubsection{\textcolor{black}{ABR algorithms}}

\textcolor{black}{
Recent research on ABR algorithms  aims to improve QoE for end users, either directly or indirectly. Direct approaches explicitly incorporate a QoE model into the 
control metric of the ABR algorithm, e.g., employing a QoE model as the reward function in an RL-based ABR algorithm. Indirect approaches focus on individual IFs, such as the bitrate, PSNR, SSIM, and VMAF to capture video quality. In addition to video quality and its stability, prominent IFs in the design and evaluation of ABR algorithms include the frequency and duration of video stalls. Efficient utilization of network bandwidth and its fair distribution among multiple sessions are common design goals. For live streaming, ABR algorithms also prioritize reducing latency. 
}

\textcolor{black}{ 
We structure our coverage of ABR designs in accordance with the taxonomy given in Section~\ref{sec:methodology-based_classification}: intuition-based (buffer-centric, throughput-centric, and hybrid), theory-based (MIP, MPC, PID, LO, BO, and other), and ML-based (RL, IL, SL, and UL), with the RL-based ABR algorithms categorized further by their reliance on A3C, A2C, AC, or other methods. Tables~\ref{tab:abr_intuition_table}, \ref{tab:abr_theory_table}, and~\ref{tab:abr_ML_table} 
summarize the intuition-based, theory-based, and ML-based ABR algorithms, respectively. In addition to the universal \textit{\textsf{core technique}} and \textit{\textsf{codec}} characteristics, each table 
describes the reviewed works with respect to: 
(1)~their application \textit{\textsf{mode}} as VoD or live streaming, (2)~\textit{\textsf{SR}} usage, (3)~employment of a \textit{\textsf{QoE model}} in design or evaluation, (4)~\textit{\textsf{bandwidth-efficiency evaluation}}, and (5)~\textit{\textsf{bandwidth-fairness evaluation}}.
}

\textcolor{black}{{\em Intuition-based methods:}}  \textcolor{black}{Laying the groundwork for buffer-centric designs, a series of \textit{\textsf{buffer-based algorithms (BBAs)}}~\cite{BBA} map the occupancy level of the playback buffer to a control metric.} BBA-0 employs piecewise linear mapping of the buffer occupancy to a bitrate. BBA-1 performs the mapping to a chunk size. BBA-2 extends BBA-1 by estimating the available network bandwidth and increasing the bitrate more aggressively during a startup phase. \textit{\textsf{Segment-aware rate adaptation (SARA)}}~\cite{sara} enhances the manifest file with chunk sizes and switches between its four adaptation modes depending on the buffer occupancy. Aiming to eliminate video stalls, the \textit{\textsf{adaptation and buffer management algorithm (ABMA+)}}~\cite{abmaplus} relies on buffer-occupancy mapping to characterize the rebuffering probability.

\input{3_Distribution_Intuition_table}

\textcolor{black}{Among throughput-centric schemes, \cite{proxypap}~caches video chunks on an \textit{\textsf{access point (AP)}} to support effective ABR streaming over the wireless link from the AP to the client.} While the AP selects chunks for prefetching into the cache, the client determines which chunks to request from either the AP or a remote server. \textit{\textsf{Low-latency prediction-based adaptation (LOLYPOP)}}~\cite{lolypop} targets live streaming and strives to improve QoE
by optimizing the operating point, a metric that combines latency, stall frequency, and bitrate-change frequency. LOLYPOP predicts TCP throughput over periods ranging from 1 to 10~s and assesses the prediction error. Interestingly, the study finds that the simple method of using the last sample as the prediction is the most accurate. Developed for streaming over mobile networks, \textit{\textsf{adaptive rate-based intelligent HTTP streaming (ARBITER+)}}~\cite{arbiterplus} addresses 
dynamic network conditions and bitrate variability through techniques such as tunable smoothing and hybrid throughput sampling. \textit{\textsf{Playback rate and priority adaptive bitrate selection (PREPARE)}}~\cite{prepare} is a throughput-centric ABR algorithm that accounts for client priority and playback speed. PREPARE improves average bitrate and stability by involving the server into prediction of the network bandwidth. Designed for live streaming, \textit{\textsf{standard low-latency video control (STALLION)}}~\cite{stallion} uses a sliding window to measure the mean and standard deviation of both bandwidth and latency. The implementation of STALLION in dash.js, a popular streaming client, outperforms the client's built-in ABR algorithm by significantly increasing the bitrate and decreasing the number of stalls.

\textcolor{black}{Representing a hybrid approach, the \textit{\textsf{fair, efficient, and stable adaptive algorithm (FESTIVE)}}~\cite{festive} combines several mechanisms to ensure efficiency, fairness, and stability in ABR streaming to multiple clients.} These mechanisms include randomized scheduling of chunk requests, harmonic-mean estimation of network bandwidth, and stateful bitrate selection with delayed updates. Pursuing similar goals, \textit{\textsf{probe and adapt (PANDA)}}~\cite{panda} incorporates estimation, smoothing, quantization, and scheduling techniques and, in particular, applies AIMD to estimate network bandwidth. Accounting for interactions between DASH and TCP, \textit{\textsf{spectrum-based quality adaptation (SQUAD)}}~\cite{squad} improves QoE by minimizing the spectrum, a metric that reflects bitrate variation. For a given ABR algorithm,  \textit{\textsf{Oboe}} computes an offline map of network conditions to an optimal configuration of algorithm parameters and automatically tunes these parameters online in response to current network conditions~\cite{akhtar2018}.  \textit{\textsf{Balancing quality of experience and traffic (BANQUET)}}~\cite{Kimura2021} aims to minimize the traffic volume while providing the QoE level specified by either the user or the streaming provider. To estimate the impact of bitrate choices on traffic and QoE, BANQUET employs brute-force search across all possible bitrate patterns for the next few chunks via predictions of buffer transitions and throughput.

\input{4_Distribution_Theory_table}

\textcolor{black}{{\em Theory-based methods:}} \textcolor{black}{\textit{\textsf{Optimized stall-cautious adaptive bitrate (OSCAR)}}~\cite{oscar} represents a MIP-based approach.} For a transient range of the buffer occupancy, it models the available network bandwidth using the Kumaraswamy distribution and formulates bitrate adaptation over a sliding look-ahead window as a \textit{\textsf{mixed-integer nonlinear programming (MINLP)}} problem. OSCAR's optimization objective combines a switching penalty with bitrate utility.

\textcolor{black}{MPC forms a prominent basis for recent ABR algorithms.
Contributing several innovations, \cite{originalmpc}~introduces two MPC-based algorithms:  \textit{\textsf{RobustMPC}} and  \textit{\textsf{FastMPC}}.} While RobustMPC performs better, FastMPC incurs significantly lower overhead. 
Additionally, this paper proposes a QoE model that underpins many subsequent ABR designs. As an MPC enhancement aimed at improving QoE, the \textit{\textsf{interest-aware approach (IAA)}}~\cite{viewinter} adjusts the bitrate by considering the user’s interest in video scenes. IAA embeds content properties into the manifest file, allowing the client to analyze these properties and quantify the user's interest in the content via the \textit{\textsf{term frequency-inverse document frequency (TF-IDF)}} method. \textit{\textsf{LDM}}~\cite{mpchttp2} utilizes MPC for live streaming and drops frames to ensure low latency. \textit{\textsf{Fugu}}~\cite{yan2020} is an MPC-based approach that predicts transmission time for each chunk via a DNN trained via SL in situ, i.e., in the actual deployment environment. To achieve low latency, \textit{\textsf{iLQR based model predictive control (iMPC)}}~\cite{towards_optimal} combines MPC with the \textit{\textsf{iterative Linear Quadratic Regulator (iLQR)}}. iMPC employs MPC to predict the available network bandwidth and iteratively linearizes the control system around its operation point to determine the bitrate via iLQR. 

\textcolor{black}{Relying on PID as its main method,
\textit{\textsf{PID-control based ABR streaming (PIA)}}~\cite{PID1} removes the \textit{\textsf{derivative (D)}} component from the standard PID controller and linearizes the closed-loop control system to maintain the buffer occupancy at a targeted level.} PIA also equips this \textit{\textsf{proportional-integral (PI)}} controller with mechanisms for faster initial ramp-up, reduction of bitrate fluctuation, and avoidance of bitrate saturation. Using the same PI controller as PIA, \textit{\textsf{quality-aware data-efficient streaming (QUAD)}}~\cite{Qaware} strives to maintain video quality at an intended level to prevent stalls, enhance playback smoothness, and reduce bandwidth consumption.

\textcolor{black}{LO-based designs include the 
\textit{\textsf{buffer occupancy based Lyapunov algorithm (BOLA)}}~\cite{bola}, which jointly optimizes playback smoothness and bitrate utility under a rate stability constraint.} BOLA provides theoretical guarantees on the achieved utility and performs excellently in practice. \textit{\textsf{Elephanta}}~\cite{Elephanta} addresses the diversity of QoE perception among different users. It offers an interface for users to adjust QoE perception parameters, models video streaming as a renewal system, and selects the bitrate by minimizing a user-specific function that combines penalties and drift.

\textcolor{black}{By employing BO, the \textit{\textsf{deep neural network for optimal tuning of adaptive video streaming controllers (ERUDITE)}}~\cite{decicco2019} configures the parameters of the \textit{\textsf{feedback linearization adaptive streaming controller (ELASTIC)}}~\cite{elastic} to jointly optimize QoE and control robustness.} At runtime, ERUDITE uses an offline-trained CNN to tune the controller parameters in accordance with real-time bandwidth measurements and video features. \cite{kimura2021a}~develops a context-aware ABR algorithm to maintain QoE at the minimum level acceptable to the user. This algorithm leverages Gaussian processes to determine the target QoE level and then selects a bitrate via BANQUET.

\textcolor{black}{Among other theory-based ABR algorithms,
\textit{\textsf{queuing theory-based rate adaptation (QUETRA)}}~\cite{quetra} uses the \textit{\textsf{Markovian deterministic single-server finite-capacity (M/D/1/K)}} queuing model to assess the buffer occupancy.} The algorithm takes into account the buffer capacity and network bandwidth, adjusting the bitrate to keep the buffer approximately half-full. QUETRA is notable for not requiring parameter tuning and performs well across various heterogeneous scenarios. To address the diversity of QoE perception among users, \textit{\textsf{affective content-aware adaptation (ACAA)}}~\cite{hu2019} considers the emotional relevance of content for different users. ACAA characterizes video chunks and users with confidence levels for six basic emotions, formulates a QoE maximization problem based on this emotional information, and solves the problem by means of~DP.

\textcolor{black}{{\em ML-based methods:}}
\textcolor{black}{
\textit{\textsf{Pensieve}}~\cite{mao2017} revolutionizes ABR streaming by applying \textit{\textsf{deep RL (DRL)}} and, in particular, the A3C method.} Pensieve formulates bitrate selection as a DRL problem and solves it using A3C where the function approximator combines \textit{\textsf{one-dimensional (1D)}} CNNs and fully connected layers. The DNN supports different encoding ladders. To accelerate state transitions, Pensieve trains its DNN with a chunk-level simulator, a technique that influences many subsequent DRL-based ABR approaches. 

\textcolor{black}{\textit{\textsf{NAS}}~\cite{Neural_content} is another A3C-based algorithm that leverages  content-aware DNNs and anytime prediction to improve QoE via SR.} For each video, the server trains multiple DNNs of different sizes and performance levels. The client picks the largest DNN able to operate in real time. Furthermore, each DNN is scalable and consists of multiple layers. This enables the client to progressively download the entire DNN, immediately benefit from the downloaded  DNN layers, and dynamically select a DNN configuration for SR of the current frames. NAS employs A3C to balance bitrate selection with progressive DNN download. \textit{\textsf{Super-resolution based adaptive video streaming (SRAVS)}}~\cite{sravs} also combines A3C with SR. Using an SR CNN~\cite{SRCNN} for video reconstruction, SRAVS maintains separate downloading and playback buffers. The separation decouples bitrate selection from reconstruction decisions, allowing for independent optimization of both processes.

\input{5_Distribution_ML_table}

\textcolor{black}{\textit{\textsf{Grad}}~\cite{liu2020} applies A3C to design ABR algorithms for SVC-encoded videos.} It mitigates SVC-related coding overhead and improves QoE through \textit{\textsf{jump-enabled hybrid coding (HYBJ)}}, where a single layer delivers multiple levels of video-quality enhancement. \cite{altamimi2020}~jointly maximizes QoE and fairness in video streaming to multiple clients over a shared bottleneck link. Its A3C actor incorporates a \textit{\textsf{long short-term memory (LSTM)}} layer, and the server dynamically configures the manifest file based on transport-layer signals  about the loss rate. With throughput measurements underlying many ABR algorithms, \textit{\textsf{accurate network throughput (ANT)}}~\cite{Yin2021} seeks to precisely model the full spectrum of available network bandwidth. ANT performs $k$-means clustering of throughput traces over short periods, trains a CNN for cluster-specific bandwidth prediction over the next period, and utilizes the prediction to select the bitrate via A3C. Also based on~A3C,  \textit{\textsf{FedABR}}~\cite{fedabr} provides faster training and preserves data privacy via \textit{\textsf{federated learning (FL)}}. After receiving from multiple clients their locally trained ABR policies, the FedABR server produces a global aggregate ABR policy and disseminates it back to the clients for further refinement of their ABR algorithms based on local data.

\textcolor{black}{\textit{\textsf{Ahaggar}}~\cite{ahaggar} trains A2C, a synchronized variant of A3C, with \textit{\textsf{distributed PPO (DPPO)}} for server-side bitrate adaptation across multiple clients.} Ahaggar leverages CMCD and CMSD for communication with clients and accelerates learning in new network conditions through meta-RL. \textcolor{black}{Additional ABR solutions in the general AC category include \textit{\textsf{Fastconv}}~\cite{fastconv}, which supports the fast training of a simple AC network by prepending an adapter that converts highly fluctuating input features into a more stable signal.} The \textit{\textsf{meta-learning framework for multi-user preferences (MLMP)}}~\cite{huo2020} utilizes multi-task DRL with PPO for policy updates, ensuring that bitrate adaptation for different users accounts for user-specific sensitivities to three QoE metrics. Designed for low-latency live streaming, the \textit{\textsf{video adaptation bitrate system (Vabis)}}~\cite{vabis} relies on \textit{\textsf{actor critic using Kronecker-factored trust region (ACKTR)}} in its server-side ABR algorithm and operates at the granularity of frames to synchronize state information during training and testing. Vabis also incorporates three playback modes on the client side and a specialized ABR regime for poor network conditions. \textit{\textsf{Stick}}~\cite{stick} combines the \textit{\textsf{deep deterministic policy gradient (DDPG)}} with BBA to improve ABR performance and reduce computational costs. Stick uses DDPG to train an AC network that controls the buffer-occupancy boundaries within the BBA approach.

\textcolor{black}{\textit{\textsf{Tiyuntsong}}~\cite{tiyuntsong} and \textit{\textsf{Ruyi}}~\cite{Ruyi} represent other RL-based ABR solutions.} Tiyuntsong employs self-play RL, where two ABR algorithms compete against each other in the same streaming environment. The rewards for the RL agents come from wins and losses in this ongoing competition, rather than from traditional QoE metrics. Additionally, each RL agent in Tiyuntsong utilizes a GAN to extract hidden features from extensive historical data. Ruyi integrates user preferences into its QoE model and leverages the model to train a \textit{\textsf{deep QL  (DQL)}} algorithm. Ruyi allows users to provide their preferences in real time, enabling adaptation to these dynamic preferences without model retraining.

\textcolor{black}{Relying on IL, \textit{\textsf{PiTree}}~\cite{PiTree} employs teacher-student learning in a simulated video player to convert DNN-based and other sophisticated ABR algorithms into accurate DT representations, thereby enabling the efficient online operation of these algorithms.} Inspired by PiTree,  \cite{reconstructing} uses DTs to reconstruct proprietary ABR algorithms in a human-interpretable manner  allowing domain experts to inspect, understand, and modify the DT representations of the algorithms. \textit{\textsf{Comyco}}~\cite{huang2020a} incorporates a solver to generate expert ABR policies aimed at maximizing QoE and trains a DNN by cloning the behavior of these expert policies. It embraces lifelong learning through continuous updates of the DNN with newly collected traces.

\textcolor{black}{\textit{\textsf{Supervised machine learning approach to adaptive video streaming over HTTP (SMASH)}}~\cite{sani2020} and \textit{\textsf{Karma}}~\cite{karma} are SL-based designs.} SMASH trains an RF classifier on outputs of nine existing ABR algorithms across various streaming scenarios, while Karma employs causal sequence modeling on a multidimensional time series and trains a GPT via SL to enhance the generalizability of ABR decisions. \textcolor{black}{Based on UL,
\textit{\textsf{Swift}}~\cite{swift}
addresses the challenges of coding overhead and latency in layered coding.} It incorporates a chain of AEs to create residual-based layered codes on the server side, a single-shot decoder on the client side, and a Pensieve-like ABR algorithm compatible with \textit{\textsf{layered neural codecs (LNCs)}}.

~\vspace{-8mm}
\subsubsection{\textcolor{black}{CDN support}} 

\textcolor{black}{
Like ABR algorithms, CDNs ultimately aim to improve QoE for end users. Recent research on CDN support focuses on achieving this goal by improving the integration of CDNs into the streaming pipeline. This includes coordinating with transcoding designs at the processing stage, collaborating with client-side ABR algorithms, deploying CDN servers, assigning users to appropriate servers, and enhancing caching performance. The proposed solutions are either specific to video streaming or also applicable to other types of traffic. Additionally, some studies investigate the utility of edge computing for video streaming.
}

\textcolor{black}{Intuition-based works include the \textit{\textsf{sequential auction mechanism (SAM)}}~\cite{crowdsourceCDN}, which operates in a crowdsourced CDN where third-party edge devices supplement CDN servers and charge CPs for leased cache space.} Another example is \textit{\textsf{intelligent network flow (INFLOW)}}~\cite{cdnselection}, an intuition-based design for dynamically selecting the most suitable CDN from multiple options. It uses measurements from video players to predict available network bandwidth and latency via LSTM. Guided by these predictions and business constraints, INFLOW intuitively selects the appropriate CDN for each player and updates the manifest file accordingly.

\textcolor{black}{Theory serves as a major foundation for CDN designs.} The \textit{\textsf{video delivery network (VDN)}}~\cite{centcontcdn} exemplifies video-specific CDN optimizations and incorporates a centralized control plane that constructs distribution trees for videos to enable scalable and highly responsive CDN operation. VDN formulates the tree construction as an integer program and approximates the program through initial solutions and early termination. To improve upon traditional CDN caching heuristics, \textit{\textsf{AdaptSize}}~\cite{adaptsize} utilizes a Markov model for content admission into the cache. To address both caching and transcoding in a radio access network, \cite{colloptim}~formulates an ILP problem to minimize CDN costs and solves it with a greedy heuristic. For enhancing ABR performance, \cite{exploryingcdncaching}~monitors video streaming of two popular CPs across three major CDNs and develops a CDN-aware variant of RobustMPC. In contrast, \textit{\textsf{FastTrack}}~\cite{fasttrack} aims to minimize the probability that stall duration in CDN-assisted video streaming exceeds a predefined threshold. FastTrack achieves this by formulating a non-convex optimization problem, dividing it into four subproblems, and solving them iteratively with an algorithm that replaces the non-convex objective function with convex approximations. To improve QoE by combining SR with edge computing, \textit{\textsf{video super-resolution and caching (VISCA)}}~\cite{edgeassisted} caches LR chunks at the edge, accounts for chunk quality and request frequency in the eviction policy, increases resolution via SR, and streams videos to players via an edge-based ABR algorithm.  

\textcolor{black}{Representing ML-based solutions, \textit{\textsf{learning-based edge with caching and prefetching (LEAP)}}~\cite{leap}  employs a DNN to prefetch and cache chunks at the edge, predicting QoE in scenarios of cache hit vs. cache miss.} Meanwhile, \textit{\textsf{RL-Cache}}~\cite{rl-cache} utilizes a feedforward neural network for cache admission and trains this network via a new DRL method that relies on direct policy search.

~
\vspace{-8mm}
\subsubsection{\textcolor{black}{QoE}}

\textcolor{black}{Since QoE models are essential for both evaluating and designing video streaming systems, research in this area heavily focuses on the interdisciplinary topic of QoE modeling. The main objective is to increase the predictive power of QoE models by applying advanced data-science techniques and incorporating new IFs, such as content characteristics and user engagement. Recent studies also emphasize personalization of QoE models to provide better service for individual users.}

\textcolor{black}{Reliance on intuition is common in QoE modeling. \textit{\textsf{YouQ}}~\cite{inferface} contributes a novel modeling technique that supports subjective tests on Facebook’s social media platform. Many ABR proposals come with intuition-based improvements to QoE models. For example, while \cite{originalmpc} introduces a QoE model that includes video quality as an IF, BOLA~\cite{bola} redefines this factor as the logarithm of the ratio between the bitrate and the lowest bitrate in the encoding ladder. Comyco~\cite{huang2020a} further changes this IF to~VMAF. In contrast, the QoE model proposed by \textit{\textsf{SENSEI}}~\cite{sensei} incorporates dynamic sensitivity to video content.}

\textcolor{black}{Recent theory-based research explores the relationship between user engagement and QoE. Whereas the queuing-theoretic analysis in \cite{ude-wiwi-12235} shows a strong correlation between these two notions,\textit{\textsf{VidHoc}}~\cite{vidhoc} utilizes user engagement as a proxy for QoE in its modeling. Specifically, VidHoc dynamically limits available network bandwidth and leverages the collected data to construct a personalized QoE model via regret minimization.}

\textcolor{black}{Among ML-based studies, \cite{porcu2019}~predicts QoE from facial expressions and gaze direction, while \cite{Laiche2021} considers DTs, RFs, and \textit{\textsf{$k$-Nearest Neighbors algorithm ($k$-NN)}} for QoE prediction based on user engagement and other factors.  \textit{\textsf{P.1203}}~\cite{p1203a1} refers to a standard QoE model that utilizes RFs to predict MOS on a five-point scale. \textit{\textsf{LSTM-QoE}}~\cite{lstmQoE} models QoE via an
LSTM network. Meanwhile, \textit{\textsf{video assessment of temporal artifacts and stalls (Video ATLAS)}}~\cite{videoatlas} expresses QoE by applying \textit{\textsf{support vector regression (SVR)}} to features related to perceptual quality, rebuffering, and memory effects. To personalize QoE models, \cite{persQoE} performs FL on sparse data and accounts for changes in IFs over time. Guided by user experience design and involving the user in a brief series of subjective assessments, \textit{\textsf{individualized QoE (iQoE)}}~\cite{iQoE} iteratively constructs an accurate personalized QoE model through active learning. Lastly, \textit{\textsf{Jade}}~\cite{humanfeed} relies on DRL with PPO to train a QoE model based on the relative ranks, rather than the absolute values, of subjective scores.}   

~
\vspace{-7mm}
\subsection{Main takeaways} \label{subsec: MT_ABRs}

Recent research on the ABR, CDN, and QoE aspects at the distribution stage primarily focuses on ABR algorithms, particularly for  the VoD streaming mode. ABR algorithms increasingly rely on ML and, especially, DRL and actor-critic methods. Studies on ABR algorithms for live streaming are less extensive, partly because the HAS paradigm offers fewer opportunities for latency reduction, which is critical for live streaming. Additionally, the adoption of DRL-based ABR algorithms in live streaming is challenging due to their high computational demands. 
For similar reasons, the use of SR at the distribution stage remains relatively rare compared to the ingestion stage. Regarding codecs, the research tendency mirrors that at the processing stage, with a predominant reliance on H.264 or H.265 as opposed to cutting-edge proprietary alternatives. However, recent work with new layered codecs shows promising results.

The general trend toward integrated designs is evident at the distribution stage, particularly in research on CDN and QoE aspects.  ABR designs that are CDN-aware or utilize well-defined QoE models become more common. Additionally, personalized QoE modeling represents an active research area. On the other hand, cooperation between the application and lower network layers struggles to gain traction. As a result, application-layer ABR algorithms primarily focus on bandwidth efficiency, while fairness in network sharing remains mostly the responsibility of the transport layer.

\vspace{-7mm}
\textcolor{black}{\section{Real-world applications}\label{sec:case_studies}}

\textcolor{black}{Commercial streaming platforms play a major role in shaping the HAS practice for long-form 2D videos. These companies inform the technical community and general public about their technologies through corporate blogs, white papers, open-source tools, standardization efforts, and academic partnerships. Additionally, researchers provide independent insights by measuring and reverse-engineering proprietary technologies.}

\vspace{-3mm}
\textcolor{black}{\subsection{Netflix}}

\textcolor{black}{Netflix regularly utilizes its technology blog to share in-depth insights into its practices and innovations. While supporting H.264, H.265, VP9, and AV1 codecs, Netflix prefers VP9 and AV1 to reduce licensing costs and enhance accessibility. As a major developer of AV1~\cite{netsatav1}, Netflix actively advocates for the codec and offers AV1 streaming on a range of devices, including television sets~\cite{netav1}. Netflix also supports AVCHi-Mobile and VP9-Mobile, which are profiles of AVC/H.264 and VP9 tailored for mobile devices~\cite{perchunk}. Additionally, Netflix employs the \textit{\textsf{dynamic optimizer (DO)}}~\cite{Netflixblog2}, a codec-agnostic system that segments video into shots and constructs representations optimized for visual perception. Aiming to enhance QoE, Netflix applies its \textit{\textsf{deep downscaler (DD)}}~\cite{deepscaler} in video preprocessing to scale down from HR to LR while preserving important visual details. DD leverages ML-based SR techniques and trains CNNs and GANs via~SL. At the distribution stage, Netflix relies on \textit{\textsf{Open Connect}}~\cite{openconnect,cdnccr},
its proprietary CDN that integrates a backbone network with tens of thousands of local servers deployed across more than one hundred countries. Developed by Netflix, VMAF~\cite{vmafcite} is an open-source technique that employs SL to fuse PSNR, SSIM, and other metrics of video quality. For Netflix and many third parties, VMAF serves as
a preferred metric for assessing video quality in applications related to ABR and QoE.
Besides, Netflix tailors a variant of VMAF for \textit{\textsf{high dynamic range (HDR)}}~\cite{vmafhdr} video, which supports a wider range of color and luminance.}

\vspace{-3mm}
\textcolor{black}{\subsection{YouTube}}

\textcolor{black}{
Similar to Netflix, YouTube supports H.264, H.265, VP9, and AV1~\cite{ytinfo}. However, YouTube focuses on VP9 as the default codec for most videos, employs H.264 for compatibility, and gradually adopts AV1, particularly for high-quality streaming~\cite{YouTubecodec,youvp9}. On average, YouTube encodes its 
VoD content into 20 representations with various combinations of bitrate and resolution. For live streaming, YouTube generates five or six representations~\cite{ytinfo} and operates in low and ultra-low latency modes, requesting chunks at intervals of 2~s and 1~s, respectively~\cite{reclive}. The distribution stage leverages the \textit{\textsf{YouTube CDN}}~\cite{youlighter}, which Google maintains specifically for serving YouTube videos. Independent measurements suggest that YouTube's proprietary ABR algorithm employs
 \textit{\textsf{quick UDP Internet connections (QUIC)}}~\cite{quicoverview} or TCP flows to download multiple chunks concurrently~\cite{requet}, utilizes less than 60\% of the available network bandwidth, sizes the playback buffer to a large duration of 80~s, and redownloads chunks in higher representations~\cite{Licc}.
}


\vspace{-3mm}
\textcolor{black}{\subsection{Amazon Prime Video}}

\textcolor{black}{
Prime Video, which is a streaming service offered by Amazon, typically supports H.264 and H.265 codecs and develops proprietary optimizations for video encoding. For example, it introduces the \textit{\textsf{encoder-aware motion compensated temporal filter (EA-MCTF)}}~\cite{EA-MCTF} for video preprocessing in conjunction with H.265 to improve video quality while maintaining low encoding time overhead. At the distribution stage, Prime Video primarily relies on \textit{\textsf{CloudFront}}, Amazon's own CDN, while also leveraging third-party CDNs to enhance performance, ensure reliability, and optimize delivery across different regions~\cite{awscdn}. Additionally, Prime Video fosters technological innovation through academic partnerships. For instance, it explores 
spatio-temporal learning of video quality~\cite{vmafprime2} and encoding parameter choices in HDR video~\cite{sportdata}. 
Similar to Netflix, Amazon Prime Video develops \textit{\textsf{ChipQA}} as a no-reference metric of video quality based on \textit{\textsf{space-time (ST) chips}}, which are localized segments of video~\cite{vmafprime1}.
}

\vspace{-3mm}
\textcolor{black}{\subsection{Twitch}}

\textcolor{black}{
Twitch primarily employs the H.264 codec with NVENC hardware acceleration~\cite{twitchguide} and advances its support for H.265~\cite{twbeta}, VP9~\cite{twitchvp9}, and AV1~\cite{twbeta}. To overcome the limitations of open-source transcoding tools, Twitch develops its own transcoder, which enhances downsampling and metadata insertion~\cite{twitchtranscoder}.
Like other major streaming platforms, Twitch relies on its own CDN for distribution~\cite{twitchcdn}, complemented by third-party CDN services. Independent measurements of Twitch's proprietary ABR algorithm indicate that it typically fills nearly 20~s of the buffer before starting playback, utilizes less than 60\% of the available network bandwidth, and assesses video quality by accounting for human perception~\cite{Licc}.
}

~
\vspace{-3mm}
\textcolor{black}{\subsection{Main takeaways}}
\textcolor{black}{
Streaming platforms such as Netflix, YouTube, Amazon Prime Video, and Twitch play a pivotal role in advancing HAS practices for long-form 2D videos. While they develop proprietary technologies for encoding, distribution, and quality assessment, they also share insights via blogs, white papers, and open-source tools. Netflix prioritizes VP9 and AV1 codecs to cut costs and improve accessibility, utilizing tools like DO, DD, and VMAF for quality optimization. YouTube focuses on VP9 and AV1, uses QUIC or TCP for efficient chunk downloads, and relies on its dedicated CDN. Amazon Prime Video employs H.264 and H.265, leverages CloudFront and third-party CDNs, and collaborates with academia on quality improvements. Twitch enhances transcoding and CDN delivery, emphasizing human perception in its ABR algorithms.
}


~
\vspace{-3mm}
\textcolor{black}{\section{Trends and future directions}\label{sec:trends_and_future}}

\textcolor{black}{
After reviewing recent research in Sections~\ref{sec:Ingestion} through~\ref{sec:Distribution} and real-world applications in Section~\ref{sec:case_studies}, we now distill current prominent trends and discuss future research directions in video streaming.
}

\vspace{-3mm}
\textcolor{black}{\subsection{Trends}\label{sec:currenttrends}}

\subsubsection{Continued growth of live streaming}

Live streaming continues to expand in traffic and attract increasing attention from researchers. At the ingestion stage, research focuses on enhancing video capture, analytics, compression, and upload to ensure low latency. The processing stage also sees some work related to live streaming, such as on-the-fly transcoding. At the distribution stage, the main research focus remains on VoD rather than live streaming.
\vspace{-5mm}
\subsubsection{Increasing diversity of devices}

The camera-equipped devices, media servers, CDN servers, and user devices that make up the streaming pipeline are diverse in type and capability. This diversity continues to grow as new devices emerge alongside legacy equipment. Ongoing changes in device capabilities enable novel pipeline configurations. For instance, smart cameras now support deep learning and play a larger role in video analytics and encoding-ladder definition, tasks traditionally handled by servers. Additionally, ABR algorithms increasingly shift from the classic client-side paradigm toward greater server-side support. Furthermore, the server infrastructure diversifies its economic models by involving CDN, edge, and cloud operators at the distribution stage. Device heterogeneity is most pronounced at both ends of the pipeline, driven by interest in new streaming modes and improvements in QoE. 
\vspace{-2mm}
\subsubsection{Integration across the end-to-end pipeline}\label{sec:integration}

Live streaming and advanced devices drive the trend toward unified solutions across the streaming pipeline, promising more efficient designs and improved end-to-end performance. For example, the distinction between initial compression in camera-equipped devices and transcoding in media servers becomes blurry, as recent designs dynamically split coding tasks between the camera-equipped device and media server to support low latency, conserve energy, decrease storage requirements, and reduce bandwidth consumption. Similarly, video analytics adopts joint designs operating at both ingestion and processing stages. SR methods are increasingly important for managing low network bandwidth during video ingestion and distribution. ABR algorithms and processing-stage tasks, such as transcoding, benefit from greater awareness of CDN, edge, and other distribution infrastructures. QoE models play a growing role in evaluating designs not only at the distribution stage, which directly interacts with end users, but also throughout the entire streaming pipeline.
\vspace{-2mm}
\subsubsection{\textcolor{black}{Shift toward ML methodologies}\label{sec:ML_trend}}

\textcolor{black}{The availability of devices with larger memory and processing capabilities also drives a greater reliance on ML methods in streaming designs. Recent results across all three stages of the streaming pipeline consistently show that ML gains popularity over intuition and theory as the basis for problem solving.
With cheaper memory and processing power, the interest in resource-intensive data-driven techniques is unsurprising. However, our survey reveals significant divergence in the ML models and training approaches employed at different stages. Ingestion-stage designs tend to rely on UL or SL with DNNs, such as CNNs. Processing-stage solutions predominantly train simpler models, such as DTs and RFs, via SL. At the distribution stage, DRL represents the most common approach, with actor-critic methods being particularly prominent.} These differences highlight challenges for the integration trend, as designing a unified ML-based solution that works effectively across all stages might be difficult.
\vspace{-7mm}
\subsubsection{\textcolor{black}{Design for better trade-offs}\label{sec:trade-offs}}

\textcolor{black}{Video streaming is a complex problem with conflicting objectives related to performance and resource consumption, making it infeasible to optimize all metrics simultaneously. Hence, practical solutions aim to offer attractive trade-offs. Technological advances impact the availability and relative costs of network bandwidth, memory, processing, energy, and other resources, thereby affecting which trade-offs are achievable.} The shift toward ML methodologies, discussed in Section~\ref{sec:ML_trend}, exemplifies new desirable trade-offs. Additionally, the integration trend broadens the range of viable trade-offs by allowing more flexible placement of functionalities across the pipeline. This search for better trade-offs is evident in the wide adoption of SR techniques, which reduce network bandwidth consumption at the cost of increased processing requirements.

~
\vspace{-8mm}
\textcolor{black}{\subsection{Future directions}}

Building on the trends discussed in Section~\ref{sec:currenttrends}, we project future developments in the field and examine their potential and challenges.

\subsubsection{\textcolor{black}{ML-based streaming}}

\textcolor{black}{Driven by increasingly affordable memory and processing power, the shift toward ML methodologies is likely to continue. Another key enabler is the wealth of unexplored opportunities, as many existing ML techniques have yet to be applied to streaming problems. For example, applying transformers to streaming deserves further investigation. Additionally, rapid advances in DNN architectures and training approaches continue to yield novel ML methods, potentially forming the basis for innovative streaming designs.} However, this abundance of research opportunities also presents challenges, particularly due to uncertainty about which directions hold the greatest promise. Specifically, as noted in Section~\ref{sec:ML_trend}, there is no clear understanding of which ML methods are most effective in supporting designs that span multiple stages of the streaming pipeline. The proliferation of ML designs across different stages also raises questions about interoperability and mutual influence. \textcolor{black}{While ML-based streaming matures, we are likely to see the development of methods tailored specifically for video streaming, rather than continued reliance on generic ML techniques.}

~
\vspace{-7.5mm}
\subsubsection{Pipeline-wide designs}

The trends of stage integration and new trade-offs, as discussed in Sections~\ref{sec:integration} and~\ref{sec:trade-offs}, converge into a future direction geared toward pipeline-wide solutions. A recent surge in research at the ingestion stage suggests a more balanced approach to all three stages and their traditional roles. Cross-stage designs now benefit from the ability to shift or split tasks between stages, optimizing resource utilization and performance. For example, moving certain analytics functions from media servers to camera-equipped devices has the potential to save network bandwidth and reduce upload latency. While pipeline-wide designs hold tremendous promise and numerous unexplored opportunities, it is desirable for unified solutions to maintain flexibility, ideally through loose coupling. SR~is likely to play a key role in these designs due to its ability to operate across all three stages of the end-to-end pipeline.

~
\vspace{-7mm}
\subsubsection{Transition to advanced codecs} 

While the surveyed research predominantly employs H.264 or H.265 due to their wide availability, a promising future direction is to build streaming systems around state-of-the-art codecs such as VVC, EVC, and LCEVC. Since cutting-edge codecs are often proprietary, research in this area is likely to involve reverse-engineering efforts, open-source initiatives, and collaborations with codec developers.

~
\vspace{-7.5mm}
\subsubsection{More ABR research with different foci} 

ABR designs vary widely in complexity and performance. Recent research often focuses on complex high-performing DNN-based algorithms, while deployed systems typically use simpler solutions of lower effectiveness. This divergence indicates the need for ABR designs with a better balance between complexity and performance. A promising direction is to improve interpretability of DNN-based ABR solutions,  leading to stronger confidence in their robustness. Although studied in other domains, work on understanding black-box ABR algorithms and converting them to simpler interpretable forms~\cite{PiTree} is relatively scarce and needs further investigation. Another promising research direction is automatic tuning of ABR algorithms. Early efforts, such as~\cite{akhtar2018,decicco2019}, explore parameter tuning via simulations. Developing efficient automatic tuning techniques for advanced DNN-based ABR algorithms represents an appealing future research area.

\vspace{-2mm}
\subsubsection{\textcolor{black}{Personalized streaming}} 

\textcolor{black}{
Despite significant variations in QoE perception among users~\cite{huo2020}, streaming services typically rely on one-size-fits-all QoE models that capture QoE as MOS, often failing to accurately reflect individual users' experiences. Personalization of QoE models shows immense promise for the enhancement of streaming services. However, inferring a user’s QoE perception non-intrusively is challenging due to the complexity of human cognition, emotions, and actions. Interdisciplinary collaborations that integrate insights from network engineering, data science, and user experience design offer significant potential for progress in this area. 
When constructing a personalized QoE model requires explicit feedback on subjective QoE perception, this feedback should be expressible, actionable, and minimal to ensure accurate QoE modeling without overburdening the user. The application of transfer learning and the development of multiple MOS-based QoE models for different reference groups, with each user assigned the most representative model, are practical alternatives. However, these methods also need to address concerns about accuracy and overhead.
}

~
\vspace{-6mm}
\subsubsection{Application-network interaction}

Video streaming within the HAS paradigm operates on top of TCP as the standard transport protocol. The independent allocation of network bandwidth by application-layer ABR logic and transport-layer CC algorithms creates problems for the efficiency, fairness, and stability of bandwidth utilization~\cite{experimental}. To address these issues, some surveyed ABR designs exploit existing transport-layer signals, while others tackle the problems by modifying the transport or network layers~\cite{minerva,sdndash}. The emergence of QUIC~\cite{quicoverview} as a promising transport protocol reinvigorates research interest in the interactions between streaming applications and underlying protocols. However, the area of application-network interaction remains underexplored, presenting opportunities for better understanding and developing integrated solutions.

~
\vspace{-6mm}
\subsubsection{Increased focus on newer modes}

While this survey covers recent developments in VoD and live modes of 2D HAS, we anticipate a growing shift in interest from VoD to live streaming. CPs, streaming platforms, and end users flock to live streaming because live content is now easy to create, profitable to distribute, and appealing to consume. From a research perspective, live streaming introduces new challenges, such as further reducing end-to-end latency within the HAS paradigm. Beyond 2D videos, 360-degree video streaming becomes increasingly important due to the wider availability of specialized equipment like omnidirectional cameras and HMDs. In addition to  360-degree video streaming, AR, VR, and MR applications, epitomized by the vision of the metaverse~\cite{metaverse}, are poised to continue attracting significant attention from both industry and research communities.

~
\vspace{-6mm}
\textcolor{black}{\subsection{Main takeaways}}
\textcolor{black}{Live streaming grows in both traffic and research interest, with a focus on reducing latency and improving the streaming pipeline. A strong trend toward integrating solutions across the end-to-end pipeline aims to improve efficiency and reduce resource consumption. The increasing diversity of devices in the streaming ecosystem drives novel configurations and encourages more server-side support in ABR algorithms. The shift toward ML-based methodologies accelerates, though challenges remain in harmonizing models across different stages of the pipeline. Future directions emphasize pipeline-wide designs, transition to advanced codecs, personalized streaming, and enhanced ABR solutions. Growing interest in application-network interaction presents opportunities to explore the integration of transport-layer signals and new protocols like QUIC. Additionally, research shifts toward newer streaming modes, particularly live streaming and immersive experiences such as 360-degree video, AR, and VR.}

~
\vspace{-3mm}
\textcolor{black}{\section{Conclusion}\label{sec:conclusion}}

\textcolor{black}{
This survey, supplemented by tutorial materials, provides a holistic overview of the end-to-end video streaming pipeline, encompassing the ingestion, processing, and distribution stages. Its focus on HAS of long-form 2D videos over CDN-assisted best-effort networks via client-side ABR algorithms reflects a dominant paradigm of modern Internet video streaming. Reviewing over 200 research papers, the survey covers key topics such as video compression, upload, transcoding, bitrate adaptation, CDN support, and QoE modeling. A new taxonomy organizes the reviewed designs by their problem-solving methodology, whether based on intuition, theory, or ML. We distinguish between MIP, MPC, PID, LO, and BO as theoretical foundations and RL, IL, SL, and UL categories of ML, with further refinement of RL into A3C, A2C, AC, and other methods. In addition, we characterize each design by its core technique and traits such as codec compatibility and SR usage. This classification and trait characterization enhance the systematic understanding of video streaming research. To connect with real-world applications, we also report on practices and innovations by major streaming platforms, such as Netflix and YouTube. The survey distills prominent current trends, including the continued growth of live streaming, shift towards ML methodologies, integration across the end-to-end pipeline, and design for better trade-offs, fueled by increasing device diversity. Looking ahead, the survey identifies promising future research directions: pipeline-wide optimization, integration of advanced codecs, further expansion of ABR research, support of personalized streaming, enhanced application-network interaction,
and stronger emphasis on newer modes of streaming. These areas represent the forefront of innovation and potential in the field.
}
\vspace{3mm}

\bibliographystyle{ACM-Reference-Format}
\bibliography{reduced_citation}

\newpage
\appendix
\section*{Appendices}
~
\vspace{-8mm}
\textcolor{black}{\section{Acronyms}\label{sec:acronyms}}
\textcolor{black}{Table~\ref{tab:acronym} lists all acronyms in alphanumeric order (left column) along with their expanded forms (right column), where capital letters match those used in the acronyms.}

\input{6_Acronyms}
\newpage


\section{Glossary}\label{sec:glossary}
\textcolor{black}{Table~\ref{tab: glossary} provides a glossary of key terms in alphanumeric order (left column) with their definitions (right column), including terms defined in the main text and those needing further explanation.}

\input{7_Glossary}

\end{document}

%% file: 1_Ingestion_table.tex
\begin{table}[t!]
\caption{\textcolor{black}{Designs at the ingestion stage of the end-to-end streaming pipeline ({\em u} abbreviates {\em unspecified}).} 
} 
\centering 
\vspace{-2mm}
\setlength{\extrarowheight}{2.13pt}

\renewcommand{\tabcolsep}{0.8mm}
\renewcommand\arraystretch{0.95}
\small
\makebox[\textwidth][c]{
\begin{tabular}{| >{\centering}m{20.8mm} | >{\centering}m{8.5mm} | >{\centering}m{7mm} | >{\centering}m{5mm} || >{\centering}m{30mm} | 
 >{\centering}m{15mm}  | >{\centering}m{3mm} |
>{\centering}m{7mm} | 
>{\centering}m{12mm} | 
>{\centering}m{10mm} | 
>{\centering\arraybackslash}m{13mm}| 
} 
\hline
{Name [reference]}  & \multicolumn{2}{c|}{Method} &  {Year} & {Core technique} & {Codec} & {SR} &    {QoE model} &  {Transport-layer signals} &  {Edge infrastructure} & {Bandwidth-efficiency evaluation}  
\\ [0.5ex]
\hline \hline

\cite{Lottermann2015}   & \multicolumn{2}{c|}{\multirow{6}{*}{Intuition}} & $2015$ & dynamic encoding ladder &  H.264  & \centering \xmark &  \centering \xmark & \centering \cmark  & \centering \xmark &  \xmark    \\ 

\cline{1-1} \cline{4-11}

\cite{Wilk2016}  & \multicolumn{2}{c|}{} & $2016$ & switch between \mbox{upload  protocols} & {\em u} & \centering \xmark &   \centering \xmark & \centering \xmark  & \centering \xmark &  \xmark     \\ 

\cline{1-1} \cline{4-11}

\cite{Park2017}   & \multicolumn{2}{c|}{} & $2017$ & AIMD-based encoding-rate control  & H.264 & \centering \xmark &  \centering \xmark &   \centering \xmark & \centering \xmark & \xmark     \\ 

\cline{1-1} \cline{4-11}

NeuroScaler~\cite{neuroscaler}  & \multicolumn{2}{c|}{} & $2022$ & zero-inference \mbox{selection of anchors}  & VP9 & \centering \cmark &    \centering \xmark & \centering \xmark  &  \centering \xmark & \xmark  \\ 

\hline
\hline

Vantage~\cite{Ray2019} & \multirow{11}{*}{Theory}& \multirow{1}{*}{\textcolor{black}{MIP}} & $2019$ & {regression heuristic} &  VP8, a VP9 predecessor  & \centering \xmark &  \centering \cmark & \centering \cmark &  \centering \xmark & \xmark    \\
\cline{1-1} \cline{3-11}

LiveSRVC~\cite{chen2021} & & \multirow{1}{*}{\textcolor{black}{MPC}} & $2021$ & SR & H.264  & \centering \cmark &   \centering \cmark & \centering \xmark & \centering \xmark & \cmark    \\

\cline{1-1} \cline{3-11}

\cite{Siekkinen2017}  & & \multirow{7}{*}{\textcolor{black}{Other}}& $2017$ & DP, greedy heuristics & SVC  & \centering \xmark &   \centering \cmark & \centering \xmark &   \centering \xmark & \xmark    \\

\cline{1-1} \cline{4-11}
CHN~\cite{Pang2019}    & &  & $2019$ & { knapsack-like problem, greedy rounding heuristic} & {\em u} & \centering \xmark  &   \centering \xmark & \centering \cmark &  \centering \cmark &  \xmark    \\

\cline{1-1} \cline{4-11}

\cite{8812206}    & & & $2019$ & relaxation-based heuristic & {\em u}  & \centering \xmark  &   \centering \cmark & \centering \xmark &  \centering \cmark &  \xmark    \\

\cline{1-1} \cline{4-11}

DDS~\cite{du2020}   & & & $2020$ & adaptive feedback control  &  H.264  & \centering \xmark &  \centering \xmark & \centering \xmark & \centering \xmark &  \cmark   \\ 

\cline{1-1} \cline{4-11}

LiveNAS~\cite{kim2020}   & & & $2020$ & {concave optimization problem, gradient ascent} & {\em u}  & \centering \cmark &     \centering \xmark & \centering \cmark &  \centering \xmark & \cmark    \\

\hline
\hline

\cite{saliencyandML}    & \multirow{8}{*}{ML}& \multirow{5}{*}{\textcolor{black}{SL}} & $2017$  & CNNs & H.265-based  & \centering \xmark &     \centering \xmark & \centering \xmark &  \xmark & \xmark  \\

\cline{1-1} \cline{4-11}

\cite{cai2020}   & & & $2020$ & CNNs & ROI-based  & \centering \xmark &   \centering \xmark & \centering \xmark & 
 \xmark &  \xmark    \\

\cline{1-1} \cline{4-11}

CrowdSR~\cite{luo2021}   & & & $2021$ & unspecified DNNs & {\em u}  & \centering \cmark &   \centering \xmark & \centering \xmark &  \centering \xmark &  \xmark    \\

\cline{1-1} \cline{4-11}

DIVA~\cite{Xu2021}  & &  & $2021$ & AlexNet variants (CNNs)  & H.264  & \centering \xmark & \centering \xmark &   \centering \xmark &  \centering \xmark &  \cmark   \\

\cline{1-1} \cline{4-11}


\textcolor{black}{MobileCodec~\cite{interframe}}   & & & \textcolor{black}{$2022$} & \textcolor{black}{CNNs}  & \textcolor{black}{MobileCodec}  & \centering \textcolor{black}{\xmark} &      \centering \textcolor{black}{\xmark} & \centering \textcolor{black}{\xmark} & \centering \textcolor{black}{\xmark} & \textcolor{black}{\xmark}    \\

\cline{1-1} \cline{3-11}

DeepFovea~\cite{kaplanyan2019} & & \multirow{3}{*}{\textcolor{black}{UL}} & $2019$ & Wasserstein GAN & DeepFovea  & \centering \xmark &  \centering \xmark &   \centering \xmark & \xmark &  \xmark \\

\cline{1-1} \cline{4-11}

Reducto~\cite{li2020a}   & & & $2020$ & $k$-means clustering  & H.264  & \centering \xmark &  \centering \xmark & \centering \xmark &  \centering \xmark & \cmark    \\

\cline{1-1} \cline{4-11}

\textcolor{black}{\cite{datascalable}}   & &  & \textcolor{black}{$2023$} & \textcolor{black}{AE}  & \textcolor{black}{data-scalable} & \centering \textcolor{black}{\xmark} &      \centering \textcolor{black}{\xmark} & \centering \textcolor{black}{\xmark} & \centering \textcolor{black}{\xmark} & \textcolor{black}{\xmark}   \\

\hline 
\end{tabular}}
\label{tab:ing_table}
\vspace{-2mm}
\end{table}

%% file: 2_Processing_table.tex

\begin{table*}[t!]
\caption{\textcolor{black}{Transcoding designs at the processing stage ({\em u} abbreviates {\em unspecified}).} 
} 
\centering 
\vspace{-2mm}
\setlength{\extrarowheight}{2.13pt}

\renewcommand{\tabcolsep}{0.5mm}
\renewcommand\arraystretch{0.95}
\small
\makebox[\textwidth][c]{
\begin{tabular}{| c | >{\centering}m{9mm} | >{\centering}m{7mm} | c || >{\centering}m{30mm} | c |c | c | c |} 
\hline
 {Name [reference]} & \multicolumn{2}{c|}{Method} & {Year} & {Core technique} & {Codec} & {Type}  &  {Performance} & {Infrastructure}  
\\ [0.5ex]
\hline \hline

\cite{f264to265} & \multicolumn{2}{c|}{\multirow{3}{*}{Intuition}} & $2017$ & statistics-driven \mbox{early termination} 
 & H.264 $\rightarrow$ H.265 & {\em u} & processing & {\em u}  \\ 

\cline{1-1} \cline{4-9}

\cite{criptotra} & \multicolumn{2}{c|}{} & $2018$  & joint crypto-transcoding & H.264, H.265 & {\em u} &  processing  & {\em u}   \\ 

\cline{1-1} \cline{4-9}

EVSO~\cite{evso} & \multicolumn{2}{c|}{} & $2018$ & frame-rate adjustment & H.264 & offline &  energy & {\em u}    \\

\hline
\hline

LwTE~\cite{lwte} & \multirow{7}{*}{Theory} & \multirow{3}{*}{\textcolor{black}{MIP}} & $2021$ & MILP, binary search & H.265 & hybrid & storage, processing & edge  \\ 

\cline{1-1} \cline{4-9}

ARTEMIS~\cite{ARTEMIS} & & & $2023$  & MILP & {\em u} & online & processing, bandwidth & CDN \\ 

\cline{1-1} \cline{4-9}

\textcolor{black}{ALPHAS~\cite{alphas}} & & & \textcolor{black}{$2025$}  & \textcolor{black}{ILP submodularity} & \textcolor{black}{H.264} & \textcolor{black}{online} & \textcolor{black}{processing} & \textcolor{black}{CDN} \\

\cline{1-1} \cline{3-9}

\cite{cdntranscoding}& & \multirow{3}{*}{\textcolor{black}{Other}} & $2015$ & Markov model  &  H.264 & hybrid & processing & CDN   \\ 

\cline{1-1} \cline{4-9}

\cite{nettrans} & & & $2018$ & context-aware \mbox{ladder optimization}  & H.264 & offline & bandwidth & {\em u} \\ 

\cline{1-1} \cline{4-9}

\cite{transenergy} & & & $2020$ & knapsack-like optimization problem & H.264 & hybrid & energy & {\em u}  \\

\hline
\hline

MAMUT~\cite{mamut} & \multirow{6}{*}{ML}& \multirow{2}{*}{\textcolor{black}{RL}} & $2018$ & multi-agent QL & H.265 & online & processing, energy & {\em u}   \\ 

\cline{1-1} \cline{4-9}

DeepLadder~\cite{Rl_ladder} & & & $2021$ & AC, dual-clip PPO, \mbox{DNN with 1D CNNs} & H.264 & online & bandwidth, storage & {\em u}   \\ 

\cline{1-1} \cline{3-9}

\cite{transratingpap} & & \multirow{4}{*}{\textcolor{black}{SL}} & $2018$ & DTs & H.265 & online & processing, energy & {\em u}  \\ 
\cline{1-1} \cline{4-9}

\cite{rftrans} & & & $2018$  & RFs & H.265 & {\em u} & processing & {\em u}  \\
\cline{1-1} \cline{4-9}

FastTTPS~\cite{fastttps} & & & $2020$ & MLP & H.264 & {\em u} & processing & {\em u} \\ 

\cline{1-1} \cline{4-9}

HEQUS~\cite{nbtrans} & & & $2021$ & NB classifiers & H.265 $\rightarrow$ VVC & {\em u} & processing & {\em u}  \\

\hline 
\end{tabular}}
\label{tab:proc_table}
\vspace{-5mm}
\end{table*}

%% file: 3_Distribution_Intuition_table.tex

\begin{table*}[t!]
\caption{\textcolor{black}{Intuition-based ABR algorithms at the distribution stage of the end-to-end streaming pipeline ({\em u} abbreviates {\em unspecified}).}} 
\vspace{-3.5mm}
\setlength{\extrarowheight}{2.13pt}
\renewcommand{\tabcolsep}{1.5mm}
\renewcommand\arraystretch{0.95}
\small
\makebox[\textwidth][c]{
\begin{tabular}{|>{\centering}m{18mm}| >{\centering}m{13.5mm} |m{5mm}||
>{\centering}m{34.5mm}|
>{\centering}m{6.5mm}|
>{\centering}m{7.6mm}|
>{\centering}m{2.6mm}|
>{\centering}m{6.3mm}|
>{\centering}m{12.1mm}| 
>{\centering\arraybackslash}m{12.1mm}|} 
\hline

Name [reference] 
& Method
& Year 
& Core technique
& Codec
& Mode
& SR 
& QoE model 
& Bandwidth efficiency evaluation 
& Bandwidth fairness evaluation \\ 
[0.5ex]  
 
\hline\hline 

BBA~\cite{BBA} & \multirow{3}{*}{\makecell{\textcolor{black}{buffer-}\\\textcolor{black}{centric}}} & $2014$ & {linear piecewise mapping} & {\em u} & VoD & \xmark & \xmark  & \xmark & \xmark \\
\cline{1-1} \cline{3-10}
SARA~\cite{sara} &  &  $2015$ & {switch between \mbox{adaptation modes}}  & {\em u} & VoD & \xmark & \xmark  & \xmark & \xmark \\
\cline{1-1} \cline{3-10}
ABMA+~\cite{abmaplus} &  & $2016$ & rebuffering-probability characterization & {\em u} & VoD & \xmark & \xmark & \cmark & \xmark  \\
\cline{1-10} 

\cite{proxypap} &  \multirow{6}{*}{\makecell{\textcolor{black}{throughput-}\\\textcolor{black}{centric}}} &  $2015$ & proxy caching  & {\em u} & VoD & \xmark & \xmark  & \xmark & \xmark \\
\cline{1-1} \cline{3-10}
LOLYPOP~\cite{lolypop} &  &  $2016$ & stall-probability prediction  & H.264 & live & \xmark & \xmark  & \xmark & \xmark  \\
\cline{1-1} \cline{3-10}
ARBITER+~\cite{arbiterplus} &  & $2018$ & hybrid throughput sampling  & H.264, H.265 & VoD & \xmark & \xmark  & \xmark & \cmark  \\
\cline{1-1} \cline{3-10}
PREPARE~\cite{prepare} &  & $2019$ &  server-client cooperation  & {\em u} & VoD & \xmark & \xmark  & \cmark & \cmark  \\
\cline{1-1} \cline{3-10}
STALLION~\cite{stallion} & & $2020$ & sliding-window measurement  & {\em u} & live & \xmark & \xmark  & \xmark & \xmark \\
\cline{1-10}

FESTIVE~\cite{festive} &   \multirow{5}{*}{\makecell{\textcolor{black}{hybrid}}}  &  $2014$ & stateful delayed bitrate update  & {\em u} & VoD & \xmark & \xmark &\cmark & \cmark  \\
\cline{1-1} \cline{3-10}
PANDA~\cite{panda} &  &  $2014$ & AIMD-based  estimation  & {\em u} & VoD & \xmark & \xmark  & \cmark & \cmark  \\
\cline{1-1} \cline{3-10}

SQUAD~\cite{squad}  &   &  $2016$& spectrum minimization  & {\em u} & VoD & \xmark & \xmark  & \xmark & \cmark  \\
\cline{1-1} \cline{3-10}

Oboe~\cite{akhtar2018} &  & $2018$  & offline parameter optimization  & {\em u}  & VoD & \xmark & \xmark  & \xmark & \xmark \\
\cline{1-1} \cline{3-10}

BANQUET~\cite{Kimura2021} &  &  $2021$ & brute-force search  & H.264 & VoD &  \xmark & \cmark  & \cmark & \xmark  \\

\hline 
\end{tabular}}
\label{tab:abr_intuition_table}
\vspace{-6mm}
\end{table*}


%% file: 4_Distribution_Theory_table.tex
\begin{table*}[t!]
\caption{\textcolor{black}{Theory-based ABR algorithms at the distribution stage of the pipeline ({\em u} abbreviates {\em unspecified}).}} 
\vspace{-2mm}
\setlength{\extrarowheight}{2.13pt}
\renewcommand{\tabcolsep}{1.5mm}
\renewcommand\arraystretch{0.95}
\small
\makebox[\textwidth][c]{
\begin{tabular}{|>{\centering}m{1.9cm}| c |m{4.8mm}||
>{\centering}m{33mm}|
>{\centering}m{10mm}|
>{\centering}m{6.8mm}|
>{\centering}m{3mm}|
>{\centering}m{7mm}|
>{\centering}m{12.1mm}| 
>{\centering\arraybackslash}m{12.1mm}|} 
\hline

Name [reference] 
& Method
& Year 
& Core technique
& Codec 
& Mode
& SR 
& QoE model 
& Bandwidth efficiency evaluation 
& Bandwidth fairness evaluation \\ 
[0.5ex]  
 
\hline\hline 
OSCAR~\cite{oscar} &  \multirow{1}{*}{\textcolor{black}{MIP}}&   $2016$ & MINLP  & H.264 & VoD & \xmark & \cmark  & \cmark & \xmark \\

\cline{1-1} \cline{2-10}
RobustMPC and FastMPC~\cite{originalmpc}  & \multirow{6}{*}{\textcolor{black}{MPC}}&  $2015$ & harmonic-mean estimation & {\em u} & VoD & \xmark & \cmark  & \xmark & \xmark \\
\cline{1-1} \cline{3-10}
IAA~\cite{viewinter} & &  $2018$ & TF-IDF  & {\em u} & VoD & \xmark & \cmark  & \xmark & \xmark \\
\cline{1-1} \cline{3-10}
LDM~\cite{mpchttp2} & &  $2020$ & frame dropping   & H.264 & live & \xmark & \cmark  & \xmark & \xmark \\
\cline{1-1} \cline{3-10}
Fugu~\cite{yan2020} & &   $2020$ & transmission-time prediction  & H.264 & VoD & \xmark & \cmark  & \xmark & \xmark \\
\cline{1-1} \cline{3-10}
iMPC~\cite{towards_optimal} & &  $2021$ & iLQR-based linearization  & H.264 & live & \xmark & \cmark  & \xmark & \xmark \\
\cline{1-1} \cline{2-10}

PIA~\cite{PID1} &  \multirow{3}{*}{\textcolor{black}{PID}}&   $2017$ & PI control with linearization   & {\em u} & VoD & \xmark & \cmark  & \xmark & \xmark \\
\cline{1-1} \cline{3-10}
QUAD~\cite{Qaware} & &  $2019$ & least-square optimization & H.264, H.265  & VoD & \xmark & \cmark  & \cmark & \xmark \\
\cline{1-1} \cline{2-10}

BOLA~\cite{bola}  &  \multirow{2}{*}{\textcolor{black}{LO}} &   $2020$ & utility maximization  & {\em u} & VoD & \xmark & \cmark  & \xmark & \xmark \\
\cline{1-1} \cline{3-10}
Elephanta~\cite{Elephanta} & &   $2020$ & renewal system    & {\em u} & VoD & \xmark & \cmark  & \xmark & \xmark \\

\cline{1-1} \cline{2-10}

ERUDITE~\cite{decicco2019} &  \multirow{3}{*}{\textcolor{black}{BO}} &  $2019$ & parameter configuration  & {\em u}  & VoD & \xmark & \cmark  & \xmark & \xmark \\
\cline{1-1} \cline{3-10}
\cite{kimura2021a} & &   $2021$ & Gaussian processes  & {\em u} & VoD & \xmark & \cmark  & \xmark & \xmark \\
\cline{1-1} \cline{2-10}

QUETRA~\cite{quetra}  &  \multirow{2}{*}{\textcolor{black}{Other}}&  $2017$ & M/D/1/K queuing  & {\em u} & VoD & \xmark & \cmark  & \xmark & \cmark \\
\cline{1-1} \cline{3-10}
ACAA~\cite{hu2019} &  &  $2019$ & DP & {\em u}  & VoD & \xmark & \cmark  & \xmark & \xmark \\

\hline 
\end{tabular}}
\vspace{-5mm}
\label{tab:abr_theory_table}
\end{table*}


%% file: 5_Distribution_ML_table.tex
\begin{table*}[t!]
\caption{\textcolor{black}{ML-based ABR algorithms at the distribution stage of the pipeline ({\em u} abbreviates {\em unspecified}).}} 
\vspace{-2mm}
\setlength{\extrarowheight}{2.13pt}
\renewcommand{\tabcolsep}{1.5mm}
\renewcommand\arraystretch{0.95}
\small
\makebox[\textwidth][c]{
\begin{tabular}{|>{\centering}m{1.9cm}| c | c |m{4.8mm}||
>{\centering}m{32mm}|
>{\centering}m{7.8mm}|
>{\centering}m{7mm}|
>{\centering}m{3mm}|
>{\centering}m{7mm}|
>{\centering}m{12.1mm}| 
>{\centering\arraybackslash}m{12.1mm}|} 
\hline

Name [reference] 
& \multicolumn{2}{c|}{Method}
& Year 
& Core technique
& Codec 
& Mode
& SR 
& QoE model 
& Bandwidth efficiency evaluation 
& Bandwidth fairness evaluation \\ 
[0.5ex]  
 
\hline\hline 
Pensieve~\cite{mao2017} & \multirow{15}{*}{\textcolor{black}{RL}}& \multirow{7}{*}{\textcolor{black}{A3C}} &  $2017$ & DNN with 1D CNNs  & {\em u} & VoD & \xmark & \cmark  & \xmark & \xmark \\
\cline{1-1} \cline{4-11}
NAS~\cite{Neural_content} & & &  $2018$ & content-aware DNNs, SR  & H.264  & VoD  & \cmark & \cmark  & \cmark & \xmark \\
\cline{1-1} \cline{4-11}
SRAVS~\cite{sravs} & & &  $2020$ & CNN, SR    & {\em u} & VoD  & \cmark & \cmark  & \xmark & \xmark \\
\cline{1-1} \cline{4-11}
Grad~\cite{liu2020}   & & &  $2020$ & DNN with 1D CNNs, HYBJ    & SVC & VoD & \xmark & \cmark  & \cmark & \xmark \\
\cline{1-1} \cline{4-11}
\cite{altamimi2020} & & &  $2020$ & LSTM, manifest update & H.264  & VoD & \xmark & \cmark  & \xmark & \cmark \\
\cline{1-1} \cline{4-11}
ANT~\cite{Yin2021} & & &  $2021$ & CNN, $k$-means clustering & {\em u} & VoD & \xmark & \cmark  & \xmark & \xmark \\
\cline{1-1} \cline{4-11}
FedABR~\cite{fedabr} & & &  $2023$ & CNN, LSTM, FL & H.264 & VoD & \xmark & \cmark  & \xmark & \xmark \\
\cline{1-1} \cline{3-11}

Ahaggar~\cite{ahaggar} & & \multirow{1}{*}{\textcolor{black}{A2C}} &  $2023$ & DPPO, DNN with 1D CNNs & H.264  & VoD & \xmark & \cmark  & \cmark & \xmark \\
\cline{1-1} \cline{3-11}

Fastconv~\cite{fastconv} & &  \multirow{4}{*}{\textcolor{black}{AC}} &  $2019$ &  CNNs   & H.264 & VoD  & \xmark & \cmark  & \xmark & \xmark \\
\cline{1-1} \cline{4-11}
MLMP~\cite{huo2020} & & &  $2020$ & PPO, LSTM   & {\em u} & VoD  & \xmark & \cmark  & \xmark & \xmark \\
\cline{1-1} \cline{4-11}
Vabis~\cite{vabis} & & &  $2020$ & ACKTR, DNNs  & {\em u} & live  & \xmark & \cmark  & \xmark & \xmark \\
\cline{1-1} \cline{4-11}
Stick~\cite{stick} & & &  $2020$ & DDPG, DNN with 1D CNNs   & H.264 & VoD  & \xmark & \cmark  & \xmark & \xmark \\
\cline{1-1} \cline{3-11}
Tiyuntsong~\cite{tiyuntsong} & & \multirow{2}{*}{\textcolor{black}{Other}}  &  $2019$ & self-play RL, GAN    & {\em u} & VoD  & \xmark & \xmark  & \xmark & \xmark \\
\cline{1-1} \cline{4-11}
Ruyi~\cite{Ruyi} & & &  $2022$ & DQL, DNN with CNNs & H.264 & VoD & \xmark & \cmark  & \xmark & \xmark \\
\cline{1-1} \cline{2-11}

PiTree~\cite{PiTree} & \multicolumn{2}{c|}{\multirow{3}{*}{\textcolor{black}{IL}}} &  $2019$ & DTs   & {\em u} & VoD  & \xmark & \xmark  & \xmark & \xmark \\
\cline{1-1} \cline{4-11}
\cite{reconstructing} & \multicolumn{2}{c|}{} &  $2020$ & DTs   & {\em u} & VoD  & \xmark & \xmark  & \xmark & \xmark \\
\cline{1-1} \cline{4-11}
Comyco~\cite{huang2020a} & \multicolumn{2}{c|}{} &  $2020$ & DNN  & H.264 & VoD & \xmark & \cmark  & \xmark & \xmark \\
\cline{1-1} \cline{2-11}

SMASH~\cite{sani2020} & \multicolumn{2}{c|}{\multirow{2}{*}{\textcolor{black}{SL}}} &  $2020$ & RFs & H.264 & VoD & \xmark & \xmark  & \xmark & \xmark \\
\cline{1-1} \cline{4-11}
Karma~\cite{karma} &\multicolumn{2}{c|}{}&  $2023$ & GPT  & H.264 & VoD & \xmark & \cmark  & \xmark & \xmark \\
\cline{1-1} \cline{2-11}

Swift~\cite{swift} & \multicolumn{2}{c|}{\multirow{1}{*}{\textcolor{black}{UL}}} &  $2022$ & AEs  & LNCs & VoD & \xmark & \cmark  & \cmark & \xmark \\

\hline 
\end{tabular}}
\label{tab:abr_ML_table}
\vspace{-4mm}
\end{table*}

%% file: 6_Acronyms.tex
\begin{table}[H]
\centering
\captionsetup{labelfont={color=black},textfont={color=black}}
\caption{\textcolor{black}{Acronyms and their expanded forms.}}
\label{tab:acronym}
\begin{tabular}{|l|p{11.35cm}|}
\hline
\textbf{Acronym} & \textbf{\hspace{4.4cm}Expanded form} \\
\hline
1D & One-Dimensional \\
2D & Two-Dimensional \\
A2C & Advantage Actor Critic \\
A3C & Asynchronous Advantage Actor Critic \\
ABMA+ & Adaptation and Buffer Management Algorithm \\
ABR & Adaptive BitRate \\
AC & Actor Critic \\
ACAA & Affective Content-Aware Adaptation \\
ACKTR & Actor Critic using Kronecker-factored Trust Region \\
AE & AutoEncoder \\
AIMD & Additive-Increase Multiplicative-Decrease \\
\textcolor{black}{ALPHAS} & \textcolor{black}{Adaptive bitrate Ladder oPtimization for multi-live HAS} \\
ANT & Accurate Network Throughput \\
AP & Access Point \\
AR & Augmented Reality \\
ARBITER+ & Adaptive Rate-Based InTElligent http stReaming \\
ARTEMIS & Adaptive bitRaTE ladder optiMIzation for live video Streaming \\
AV1 & Aomedia Video 1 \\
AVC & Advanced Video Coding \\
B-frame & Bipredictive frame \\
BANQUET & BAlaNcing QUality of Experience and Traffic  \\
BBA & Buffer-Based Algorithm \\
BBR & Bottleneck Bandwidth and Round-trip propagation time \\
BO & Bayesian Optimization \\
BOLA & Buffer Occupancy based Lyapunov Algorithm \\
CC & Congestion Control \\
CDN & Content Delivery Network \\
CHN & Content Harvest Network \\
CMCD & Common Media Client Data \\
CMSD & Common Media Server Data \\
CNN & Convolution Neural Network \\
CP & Content Provider \\
CTU & Coding Tree Unit \\

\hline
\end{tabular}
\end{table}
\newpage

\begin{table}[H]
\centering
\captionsetup{labelfont={color=black},textfont={color=black}}
\caption*{\textcolor{black}{Table 6. Acronyms and their expanded forms (continued).}}
\label{tab:acronym_continued}
\begin{tabular}{|l|p{11.35cm}|}
\hline
\textbf{Acronym} & \textbf{\hspace{4.4cm}Expanded form} \\
\hline
CU & Coding Unit \\
D & Derivative \\
DASH & Dynamic Adaptive Streaming over Http \\
DCT & Discrete Cosine Transform \\
DD & Deep Downscaler \\
DDPG & Deep Deterministic Policy Gradient \\
DDS & Dnn-Driven Streaming \\
DNN & Deep Neural Network \\
DO & Dynamic Optimizer \\
DP & Dynamic Programming \\
DPPO & Distributed Proximal Policy Optimization \\
DQL & Deep Q-Learning \\
DRL & Deep Reinforcement Learning \\
DST & Discrete Sine Transform \\
DT & Decision Tree \\
EA-MCTF & Encoder-Aware Motion Compensated Temporal Filter \\
ELASTIC & fEedback Linearization Adaptive STreamIng Controller \\
ERUDITE & dEep neuRal network for optimal tUning of aDaptive vIdeo sTreaming controllErs \\
EVC & Essential Video Coding \\
EVSO & Environment-aware Video Streaming Optimization \\
FESTIVE & Fair, Efficient, and Stable adapTIVE algorithm \\
FL & Federated Learning \\
FastTTPS & Fast video Transcoding Time Prediction and Scheduling \\
fps & frames per second \\
GAN & Generative Adversarial Network \\
GCC & Google Congestion Control \\
GOP & Group Of Pictures \\
GPT & Generative Pre-trained Transformer \\
GPU & Graphics Processing Unit \\
HAS & Http Adaptive Streaming \\
HDR & High Dynamic Range \\
HEQUS & HEvc-based QUadtree Splitting \\
HEVC & High Efficiency Video Coding \\
HLS & Http Live Streaming \\
HMD & Head-Mounted Display \\
HR & High-Resolution \\
HTTP & HyperText Transfer Protocol \\
HYBJ & Jump-enabled HYBrid coding \\
\hline
\end{tabular}
\end{table}
\newpage

\begin{table}[H]
\centering
\captionsetup{labelfont={color=black},textfont={color=black}}
\caption*{\textcolor{black}{Table 6. Acronyms and their expanded forms (continued).}}
\label{tab:acronym_continued}
\begin{tabular}{|l|p{11.35cm}|}
\hline
\textbf{Acronym} & \textbf{\hspace{4.4cm}Expanded form} \\
\hline
I-frame & Intra-frame \\
IAA & Interest-Aware Approach \\
IF & Influence Factor \\
IL & Imitation Learning \\
ILP & Integer Linear Programming \\
INFLOW & Intelligent Network FLOW \\ 
ISP & Internet Service Provider \\
iLQR & iterative Linear Quadratic Regulator \\
iMPC & ilqr based Model Predictive Control \\
iQoE & individualized Quality of Experience \\
LCEVC & Low Complexity Enhancement Video Coding \\
LEAP & Learning-based Edge with cAching and Prefetching \\
LNC & Layered Neural Codec \\
LO & Lyapunov Optimization \\
LOLYPOP & LOw-LatencY PredictiOn-based adaPtation \\ 
LR & Low-Resolution \\
LSTM & Long Short-Term Memory \\
LwTE & Light-weight Transcoding at the Edge \\
$k$-NN & $k$-Nearest Neighbors algorithm \\
M/D/1/K & Markovian Deterministic Single-server finite-Capacity\\
MILP & Mixed-Integer Linear Programming \\
MINLP & Mixed-Integer NonLinear Programming \\
MIP & Mixed-Integer Programming \\
ML & Machine Learning \\
MLMP & Meta-Learning framework for Multi-user Preferences \\
MLP & MultiLayer Perceptron \\
MOS & Mean Opinion Score \\
MPC & Model Predictive Control \\
MPEG & Moving Picture Experts Group \\
MR & Mixed Reality \\
NADA & Network-Assisted Dynamic Adaptation\\
NB & Naive Bayes \\
NDN & Named Data Networking \\
NP &  Nondeterministic Polynomial time \\
NVENC & NVidia ENCoder \\
OSCAR & Optimized Stall-Cautious Adaptive bitRate \\
P-frame & Predictive frame \\
P2P & Peer-To-Peer \\
\hline
\end{tabular}
\end{table}

\begin{table}[H]
\centering
\captionsetup{labelfont={color=black},textfont={color=black}}
\caption*{\textcolor{black}{Table 6. Acronyms and their expanded forms (continued).}}
\label{tab:acronym_continued}
\begin{tabular}{|l|p{11.35cm}|}
\hline
\textbf{Acronym} & \textbf{\hspace{4.4cm}Expanded form} \\
\hline
PANDA & Probe AND Adapt \\
PI & Proportional-Integral \\
PIA & PId-control based Abr streaming \\
PID & Proportional-Integral-Derivative \\
PPO & Proximal Policy Optimization \\
PREPARE & Playback RatE and Priority Adaptive bitRatE selection \\
PSNR & Peak Signal-to-Noise Ratio \\
QL & Q-Learning \\
QP & Quantization Parameter \\
QT & QuadTree \\
QUAD & QUality-Aware Data-efficient streaming \\
QUETRA & QUEuing Theory-based Rate Adaptation \\
QUIC & Quick Udp Internet Connections \\
QoE & Quality of Experience \\
QoS & Quality of Service \\
RF & Random Forest \\
RL & Reinforcement Learning \\
ROI & Region Of Interest \\
RTMP & Real-Time Messaging Protocol \\
SAM & Sequential Auction Mechanism \\
SARA & Segment-Aware Rate Adaptation \\
SCReAM & Self-Clocked Rate Adaptation for Multimedia \\
SDN & Software-Defined Networking \\
SL & Supervised Learning \\
SMASH & Supervised Machine learning Approach to adaptive video Streaming over Http \\
SQUAD & Spectrum-based QUality ADaptation \\ 
SR & Super Resolution \\
SRAVS & Super-Resolution based Adaptive Video Streaming \\
SSIM & Structural Similarity Index Measure \\
SSR & Short-form video Streaming and Recommendation \\
ST & Space-Time \\
STALLION & STAndard Low-Latency vIdeo cONtrol \\
SVC & Scalable Video Coding \\
SVR & Support Vector Regression \\
TCP & Transmission Control Protocol \\
TF-IDF & Term Frequency-Inverse Document Frequency \\
UDP & User Datagram Protocol \\
UL & Unsupervised Learning \\
\hline
\end{tabular}
\end{table}

\begin{table}[H]
\centering
\captionsetup{labelfont={color=black},textfont={color=black}}
\caption*{\textcolor{black}{Table 6. Acronyms and their expanded forms (continued).}}
\label{tab:acronym_continued}
\begin{tabular}{|l|p{11.35cm}|}
\hline
\textbf{Acronym} & \textbf{\hspace{4.4cm}Expanded form} \\
\hline
VCE & Video Coding Engine \\
VDN & Video Delivery Network \\
VISCA & VIdeo Super-resolution and CAching \\
VMAF & Video Multimethod Assessment Fusion \\
VR & Virtual Reality \\
VVC & Versatile Video Coding \\
Vabis & Video adaptation bitrate system \\
Video ATLAS & Video Assessment of TemporaL Artifacts and Stalls \\
VoD & Video on Demand \\
WebRTC & Web Real-Time Communication \\
\hline
\end{tabular}
\end{table}

%% file: 7_Glossary.tex
\begin{table}[H]
\centering
\captionsetup{labelfont={color=black},textfont={color=black}}
\caption{\textcolor{black}{Glossary.}}
\label{tab: glossary}
\begin{tabular}{|l|p{10.4cm}|}
\hline
\textcolor{black}{\hspace{0.7cm}\textbf{Name}} & \textcolor{black}{\hspace{4.4cm}\textbf{Definition}} \\
\hline
\multirow{2}{*}{\textcolor{black}{360-degree video}}&\textcolor{black}{A video format that immerses users in a panoramic environment, enabling them to look around in all directions.}\\

\hline
\multirow{4}{*}{\textcolor{black}{A2C}}&\textcolor{black}{An on-policy model-free RL algorithm that extends AC by incorporating an advantage function to reduce variance in the critic’s value function estimate. It updates both actor and critic simultaneously, using the advantage function to guide the actor toward better action choices.}\\

\hline
\multirow{4}{*}{\textcolor{black}{A3C}}&\textcolor{black}{ An on-policy model-free RL algorithm that uses multiple agents running in parallel in different environments to asynchronously update a global model. Each agent computes an advantage estimate, allowing the algorithm to stabilize learning and improve scalability over standard A2C.}\\
\hline
\multirow{4}{*}{\textcolor{black}{ABR}}&\textcolor{black}{ A streaming technique where the system divides the video into chunks and encodes each at different size-quality levels. The player dynamically selects the appropriate chunk during playback to ensure smooth streaming, minimize buffering, and optimize the user experience across varying network conditions.}\\
\hline
\multirow{4}{*}{\textcolor{black}{AC}}&\textcolor{black}{ An on-policy model-free RL algorithm where the actor selects actions based on the current policy, and the critic evaluates those actions by estimating the value function. The system updates both components simultaneously to improve the policy and value function.}\\
\hline
\multirow{4}{*}{\textcolor{black}{ACKTR}}&\textcolor{black}{ An on-policy model-free RL algorithm that extends AC by incorporating a trust region method with Kronecker-factored approximations of the Fisher information matrix. This method optimizes policy updates more efficiently by controlling the step size, improving stability, and enhancing convergence.}\\
\hline
\multirow{4}{*}{\textcolor{black}{AE}}&\textcolor{black}{ A type of neural network for UL that encodes input data into a lower-dimensional representation and then reconstructs it back to its original form, commonly used for dimensionality reduction, feature learning, and noise reduction.}\\
\hline
\multirow{4}{*}{\textcolor{black}{AIMD}}&\textcolor{black}{ A CC algorithm used in computer networks that gradually increases the data transmission rate (additive increase) while not detecting congestion and sharply reduces it (multiplicative decrease) upon detecting congestion, helping stabilize network throughput and avoid overload.}\\
\hline
\multirow{3}{*}{\textcolor{black}{AR}}&\textcolor{black}{ An immersive media experience that overlays digital elements on top of the real-world view, enhancing the user's interaction with their physical environment through video content.}\\
\hline
\end{tabular}
\end{table}

\begin{table}[H]
\centering
\captionsetup{labelfont={color=black},textfont={color=black}}
\caption*{Table 7. Glossary (continued).}
\begin{tabular}{|l|p{10.4cm}|}
\hline
{\hspace{0.7cm}\textbf{Name}} & \textcolor{black}{\hspace{4.4cm}\textbf{Definition}} \\
\hline
\multirow{3}{*}{\textcolor{black}{BO}}&\textcolor{black}{ An optimization method designed for black-box functions with expensive evaluations. It uses a probabilistic model, typically a Gaussian process, to predict the function’s behavior. }\\
\hline
\multirow{4}{*}{\textcolor{black}{Bitrate}}&\textcolor{black}{ The amount of data processed in a video stream per unit of time, typically measured in kilobits or megabits per second (Kbps or Mbps). It directly influences both video quality and data consumption, with higher bitrates generally offering better quality at the cost of increased data usage.}\\
\hline
\multirow{3}{*}{\textcolor{black}{CDN}}&\textcolor{black}{ A system of cache servers distributed across wide geographical areas to improve the performance of content delivery from CPs to end users by providing data closer to the final user.}\\
\hline
\multirow{3}{*}{\textcolor{black}{CNN}}&\textcolor{black}{ A type of DNN designed specifically for visual data processing. It leverages convolutional layers to extract spatial features from input data, pooling layers to reduce dimensionality, and fully connected layers for classification or regression.}\\
\hline
\multirow{2}{*}{\textcolor{black}{Chunk}}&\textcolor{black}{ A small self-contained segment of a video, encoded at specific resolution and bitrate settings, allowing independent download and playback.}\\
\hline
\multirow{3}{*}{\textcolor{black}{Codec}}&\textcolor{black}{ A hardware or software tool that compresses and decompresses digital media by applying spatial and temporal compression, reducing data size and enabling efficient storage and transmission of content.}\\
\hline
\multirow{3}{*}{\textcolor{black}{Congestion control}}&\textcolor{black}{ A set of techniques used to manage and mitigate network congestion, ensuring efficient data flow and preventing network bottlenecks. It dynamically adjusts transmission rates based on real-time network conditions.}\\
\hline
\multirow{4}{*}{\textcolor{black}{DDPG}}&\textcolor{black}{ An off-policy model-free AC-based RL algorithm, designed for environments with continuous action spaces. It uses a deterministic policy and employs a target network along with experience replay to stabilize learning, making it suitable for complex control tasks.}\\
\hline
\multirow{3}{*}{\textcolor{black}{DCT}}&\textcolor{black}{ A mathematical transformation in video compression that converts spatial data into frequency components, allowing less important data to remove. This reduces data size while maintaining visual quality.}\\

\hline
\multirow{2}{*}{\textcolor{black}{DNN}}&\textcolor{black}{ A type of neural network with multiple hidden layers between the input and output, enabling it to model complex patterns and relationships in data.}\\
\hline
\multirow{3}{*}{\textcolor{black}{DP}}&\textcolor{black}{ An optimization method that solves complex problems by breaking them into smaller overlapping subproblems, solving each subproblem only once, and storing the results to avoid redundant computations.}\\
\hline
\multirow{4}{*}{\textcolor{black}{DT}}&\textcolor{black}{ An ML algorithm used for SL classification and regression tasks. It models decisions and their possible consequences as a tree structure, where each internal node represents a decision based on a feature, each branch represents the outcome of that decision, and each leaf node represents the final prediction or class label.}\\
\hline
\end{tabular}
\end{table}

\begin{table}[H]
\centering
\captionsetup{labelfont={color=black},textfont={color=black}}
\caption*{Table 7. Glossary (continued).}
\begin{tabular}{|l|p{10.4cm}|}
\hline
{\hspace{0.7cm}\textbf{Name}} & \textcolor{black}{\hspace{4.4cm}\textbf{Definition}} \\
\hline
\multirow{2}{*}{\textcolor{black}{Encoding ladder}}&\textcolor{black}{ A predefined set of resolutions and bitrates used for encoding a video into multiple versions to balance bandwidth consumption and user experience.}\\
\hline
\multirow{3}{*}{\textcolor{black}{FL}}&\textcolor{black}{A decentralized ML approach where multiple devices or systems collaborate to train a model while keeping their data local. It enables model training without sharing sensitive data, ensuring privacy and security.}\\
\hline
\multirow{4}{*}{\textcolor{black}{Frame rate}}&\textcolor{black}{The number of frames displayed in a video stream per unit of time, typically measured in frames per second (fps). It directly affects motion smoothness and visual fluidity, with higher frame rates generally providing smoother playback at the cost of increased processing and bandwidth requirements.}\\
\hline
\multirow{2}{*}{\textcolor{black}{Frames}}&\textcolor{black}{ Individual still images in a video sequence that, when displayed in rapid succession, create the illusion of motion.}\\
\hline
\multirow{5}{*}{\textcolor{black}{GAN}}&\textcolor{black}{ A type of neural network consisting of two components: a generator and a discriminator. The generator creates synthetic data, while the discriminator evaluates its authenticity by comparing it to real data, with both networks competing to improve their performance. This adversarial process results in the generation of realistic synthetic data.}\\
\hline
\multirow{2}{*}{\textcolor{black}{GOP}}&\textcolor{black}{ A sequence of video frames that starts with an I-frame, followed by dependent frames like P-frames and B-frames.}\\
\hline
\multirow{4}{*}{\textcolor{black}{Gaussian processes}}&\textcolor{black}{ An ML algorithm used for SL tasks that models data using a distribution over functions. It uses a kernel function to define covariance between data points and makes predictions by calculating a distribution over possible outputs, providing both a mean and uncertainty estimate.}\\
\hline
\multirow{2}{*}{\textcolor{black}{Gradient ascent}}&\textcolor{black}{ An optimization method used to maximize a function by iteratively adjusting its parameters in the direction of the steepest increase of the function.}\\
\hline
\multirow{2}{*}{\textcolor{black}{I-frame}}&\textcolor{black}{ A type of video frame that functions independently of other frames and serves as a reference for decoding other frames within a GOP.}\\
\hline
\multirow{2}{*}{\textcolor{black}{IL}}&\textcolor{black}{ An ML approach where a model learns to perform tasks by observing and mimicking expert demonstrations.}\\
\hline
\multirow{3}{*}{\textcolor{black}{In-transit computing}}&\textcolor{black}{ An approach to processing data while it transits across a network, typically at intermediary points such as edge devices or network nodes, instead of waiting for it to reach its final destination. }\\
\hline
\multirow{4}{*}{\textcolor{black}{$k$-means clustering}}&\textcolor{black}{ An ML algorithm used for UL that partitions data into $k$ distinct clusters based on feature similarity. It assigns data points to the nearest centroid, updates the centroids, and repeats the process until convergence, aiming to minimize the variance within each cluster.}\\
\hline
\multirow{2}{*}{\textcolor{black}{LQR}}&\textcolor{black}{ A control algorithm used to determine the optimal control inputs for a linear dynamic system by minimizing a quadratic cost function.}\\
\hline
\end{tabular}
\end{table}

\begin{table}[H]
\centering
\captionsetup{labelfont={color=black},textfont={color=black}}
\caption*{Table 7. Glossary (continued).}
\begin{tabular}{|l|p{10.4cm}|}
\hline
{\hspace{0.7cm}\textbf{Name}} & \textcolor{black}{\hspace{4.4cm}\textbf{Definition}} \\
\hline
\multirow{3}{*}{\textcolor{black}{LSTM}}&\textcolor{black}{ A type of DNN designed to handle long-term dependencies in sequential data. It uses memory units with gates to regulate the flow of information, allowing the model to retain important data over time.}\\
\hline
\multirow{3}{*}{\textcolor{black}{MINLP}}&\textcolor{black}{ A modeling approach used to represent problems that involve both nonlinear relationships and mixed variable types, including continuous and discrete (integer) variables.}\\
\hline
\multirow{3}{*}{\textcolor{black}{MIP}}&\textcolor{black}{ A modeling approach used to represent problems that involve both linear relationships and mixed variable types, including continuous and discrete (integer) variables.}\\
\hline
\multirow{3}{*}{\textcolor{black}{MLP}}&\textcolor{black}{ A type of feedforward neural network consisting of an input layer, one or more hidden layers, and an output layer. Each neuron in one layer connects fully to neurons in the subsequent layer.}\\
\hline
\multirow{3}{*}{\textcolor{black}{MPC}}&\textcolor{black}{ A control algorithm that uses a system model to predict future states and optimize control inputs over a finite time horizon. It solves an optimization problem at each time step, taking into account system dynamics, constraints, and desired outcomes.}\\
\hline
\multirow{2}{*}{\textcolor{black}{MR}}&\textcolor{black}{A hybrid immersive media experience combining elements of both AR and VR, where virtual and real-world objects coexist and interact in real time.}\\
\hline
\multirow{3}{*}{\textcolor{black}{Manifest}}&\textcolor{black}{ A description of available video chunks, their bitrates, resolutions, playback order, and other metadata such as codec information, segment duration, and subtitle tracks.}\\
\hline
\multirow{2}{*}{\textcolor{black}{P-frame}}&\textcolor{black}{ A type of video frame encoded by referencing previous frames, storing only the differences to reduce data size while maintaining quality.}\\
\hline
\multirow{4}{*}{\textcolor{black}{PID}}&\textcolor{black}{ A control algorithm that adjusts the control input based on three terms: proportional, integral, and derivative, which help minimize the error between a desired setpoint and the measured value by considering the current error, accumulated past error, and rate of change of the error. }\\
\hline
\multirow{3}{*}{\textcolor{black}{PPO}}&\textcolor{black}{ An on-policy model-free RL algorithm that improves policy updates by limiting the magnitude of changes using a surrogate objective function. PPO strikes a balance between performance and stability.}\\
\hline
\multirow{5}{*}{\textcolor{black}{PSNR}}&\textcolor{black}{ An image quality metric that compares the original and compressed images pixel by pixel by evaluating the ratio between the maximum possible signal and the noise introduced. The metric assesses the mean squared error between the two images. Higher PSNR values suggest better image quality, as less distortion has occurred during compression.}\\
\hline
\multirow{3}{*}{\textcolor{black}{QL}}&\textcolor{black}{ An off-policy model-free RL algorithm that learns the value of state-action pairs using a Q-function. The algorithm updates the Q-values iteratively based on expected future rewards. QL extends to continuous action spaces by using DNNs.}\\
\hline
\end{tabular}
\end{table}

\begin{table}[H]
\centering
\captionsetup{labelfont={color=black},textfont={color=black}}
\caption*{Table 7. Glossary (continued).}
\begin{tabular}{|l|p{10.4cm}|}
\hline
{\hspace{0.7cm}\textbf{Name}} & \textcolor{black}{\hspace{4.4cm}\textbf{Definition}} \\
\hline
\multirow{4}{*}{\textcolor{black}{QoE}}&\textcolor{black}{ A subjective measure that captures how satisfied a user is with a service, based on individual perceptions and experiences. It evaluates factors which contribute to the user's personal judgment. Unlike objective metrics that focus on technical parameters, QoE emphasizes the human perspective.}\\
\hline
\multirow{4}{*}{\textcolor{black}{QoS}}&\textcolor{black}{ An objective measure of the performance of a network or service, focusing on measurable factors such as bandwidth, latency, jitter, and packet loss. Unlike QoE, which is subjective, QoS quantifies the technical performance of the system, ensuring that these parameters meet predefined thresholds.}\\
\hline
\multirow{3}{*}{\textcolor{black}{RF}}&\textcolor{black}{ An ML algorithm used for SL that constructs an ensemble of decision trees, each trained on a random subset of the data. The algorithm aggregates the outputs of individual trees to make the final prediction.}\\
\hline
\multirow{3}{*}{\textcolor{black}{RL}}&\textcolor{black}{ An ML approach where an agent learns to make decisions through interactions with an environment, aiming to maximize cumulative rewards over time by exploring and exploiting different actions.}\\
\hline
\multirow{4}{*}{\textcolor{black}{ROI}}&\textcolor{black}{ A selected area within an image or video where analysis, processing, or encoding focuses. In encoding, this region receives higher priority to enhance its quality, while other less important areas compress more aggressively, optimizing both visual quality and compression efficiency.}\\
\hline
\multirow{3}{*}{\textcolor{black}{Rebuffering}}&\textcolor{black}{ A stall caused by the depletion of the buffer, often due to insufficient data delivered to maintain continuous playback, typically resulting from poor network conditions or inadequate buffer size. }\\
\hline
\multirow{2}{*}{\textcolor{black}{Resolution}}&\textcolor{black}{ The number of pixels in each dimension that a video frame contains, determining its visual detail and quality.}\\
\hline
\multirow{3}{*}{\textcolor{black}{SL}}&\textcolor{black}{ An ML approach where a model learns from labeled data, with each input paired with a corresponding ground truth or correct output, enabling the model to make predictions or classifications based on these learned associations.}\\
\hline
\multirow{2}{*}{\textcolor{black}{SR}}&\textcolor{black}{ A technique in image and video processing to enhance the resolution of a video or image beyond its original quality.}\\
\hline
\multirow{5}{*}{\textcolor{black}{SSIM}}&\textcolor{black}{ An image quality metric that measures the similarity between two images by comparing luminance, contrast, and structural information. It calculates local patterns of pixel intensities using a sliding window. The metric considers the image’s structural components and human visual perception more effectively than more traditional metrics like PSNR.}\\
\hline
\multirow{4}{*}{\textcolor{black}{SVR}}&\textcolor{black}{ An ML algorithm used for SL regression tasks. It finds a hyperplane that best fits the data while allowing for a margin of error and predicts new data points based on this hyperplane. SVR is particularly effective in situations with nonlinear relationships between variables and handles high-dimensional feature spaces.}\\
\hline
\end{tabular}
\end{table}

\begin{table}[H]
\centering
\captionsetup{labelfont={color=black},textfont={color=black}}
\caption*{Table 7. Glossary (continued).}
\begin{tabular}{|l|p{10.4cm}|}
\hline
{\hspace{0.7cm}\textbf{Name}} & \textcolor{black}{\hspace{4.4cm}\textbf{Definition}} \\
\hline
\multirow{2}{*}{\textcolor{black}{Short-form videos}}&\textcolor{black}{ Video content typically under one minute long, optimized for quick consumption and highly popular on social media platforms.}\\
\hline
\multirow{3}{*}{\textcolor{black}{Stall}}&\textcolor{black}{ Any interruption or pause in video playback from various causes, such as buffer depletion, processing delays, or decoding errors. A stall indicates that playback is unable to continue seamlessly, regardless of the specific reason behind it.}\\

\hline
\multirow{4}{*}{\textcolor{black}{TF-IDF}}&\textcolor{black}{ A statistical measure used to assess the importance of a word within a document relative to a collection of documents. It balances the frequency of the word in the document (TF) with how rare the word is across the entire corpus (IDF), emphasizing terms that are unique and relevant to a specific document.}\\
\hline
\multirow{4}{*}{\textcolor{black}{Transcoding}}&\textcolor{black}{ The process of converting a video file from one codec or format to another to ensure compatibility with various devices or platforms. Unlike encoding, which compresses raw video data into a specific format, transcoding reformats an already encoded file, typically for optimization, quality adjustment, or compatibility.}\\
\hline
\multirow{3}{*}{\textcolor{black}{UL}}&\textcolor{black}{ An ML approach where the model identifies patterns or structures in unlabeled data without predefined outcomes, aiming to discover underlying relationships or groupings within the data.}\\

\hline
\multirow{3}{*}{\textcolor{black}{VMAF}}&\textcolor{black}{A video quality metric developed by Netflix that predicts perceived video quality by combining multiple quality assessment methods. It uses an SVR model to learn the optimal weights for different quality metrics based on human perception.}\\

\hline
\multirow{3}{*}{\textcolor{black}{VR}}&\textcolor{black}{A fully immersive media experience where the viewer immerses in a virtual environment, often using specialized headsets, creating a sense of being inside the content.}\\
\hline
\end{tabular}
\end{table}